\documentclass[aps, prb, twocolumn,amsmath,amssymb,floatfix, reprint, longbibliography]{revtex4-2}

\usepackage{graphicx}
\usepackage{svg}
\usepackage{times}
\usepackage{dcolumn}
\usepackage{hyperref}
\usepackage{physics}
\usepackage{siunitx}
\hypersetup{colorlinks, allcolors=blue}
\usepackage{lipsum}
\usepackage{comment}
\usepackage[labelfont=rm]{caption}
\usepackage{subcaption}
\usepackage{gensymb}
\usepackage{soul}
\usepackage{chemformula}

\newcommand{\kk}{{\mathbf{k}}}
\newcommand{\nn}{{\mathbf{n}}}

\newcommand{\rr}{{\mathbf{r}}}

\newcommand{\bq}{{\mathbf{q}}}
\newcommand{\vF}{v_{\mathrm{F}}}
\newcommand{\e}[1]{\mathrm{e}^{#1}}
\newcommand{\bsigma}{{\boldsymbol{\sigma}}}
\newcommand{\kF}{{k_{\text{F}}}}

\newcommand{\bhat}[1]{\hat{\mathbf{#1}}}

\DeclareMathOperator{\sgn}{sgn}

\newcommand{\waschanged}[1]{{#1}}

\begin{document}
\title{Exchange tensors, generalized RKKY interactions, and magnetization dynamics in heterostructures of ferromagnets and topological insulators}

\author{Christian Svingen Johnsen and Asle Sudb\o}
	\affiliation{Center for Quantum Spintronics, Department of Physics, Norwegian University of Science and Technology, NO-7491 Trondheim, Norway}

\begin{abstract}

We present a comprehensive theoretical analysis of magnetic heterostructures composed of ferromagnetic (FM) layers interfaced with three-dimensional topological insulators (TIs). Integrating out the topological surface states and computing the spin determinant to second order in spins, we derive the effective generalized Ruderman–Kittel–Kasuya–Yosida (RKKY) exchange interactions mediated by topological surface states. These interactions inherently incorporate spin-momentum locking and anisotropic spin susceptibilities stemming from the Dirac-like dispersion of the TI surface electrons. The analysis reveals that the interplay between the spin-orbit coupling intrinsic to the TI and the magnetization in the FM layer induces highly nonlocal and retarded, chiral, and Dzyaloshinskii-Moriya (DM)-like contributions to the effective spin Hamiltonian. Furthermore, the spin dynamics is studied through a derivation of the LLG equation for this problem. The induced interactions renormalize many of the FM's intrinsic properties, but a \waschanged{new} term in the LLG equation is induced that is related to the rate of change of the magnetization's curl, which is relevant to skyrmion dynamics. The magnon dispersion exhibits modifications due to the TI-mediated interactions, including a \waschanged{softened inertial spin-wave mode} and tunable magnon gaps, sensitive to a tunable chemical potential and interfacial exchange coupling strength. The results also apply to finite temperatures. They elucidate topologically induced magnetic phenomena and pave the way for engineering exotic spin textures, such as skyrmions and chiral domain walls, in TI-FM hybrid systems with tunable interactions.

\end{abstract}
\maketitle

\section{Introduction}

\label{sec:Intro}

Topological defects, such as vortices and domain walls, arise from broken symmetries in condensed matter systems and play key roles in phase transitions and material properties. Skyrmions, stable spin textures with topological protection, are particularly important for their potential in low-energy spintronic devices, offering robustness, scalability, and efficient information storage.
The existence of magnetic skyrmions has been considered possible at least since 2006 \cite{Rler2006}. These highly particular windings of the magnetization field in magnets are topologically protected magnetic textures which can arise from various mechanisms and in various materials, yielding different types of skyrmions \cite{Finocchio2016}. They have been observed in MnSi \cite{Muhlbauer2009, Neubauer2009, Pappas2009} and various other materials \cite{Yu2010, Yu2010-2, Seki2012}, including antiferromagnets (AFMs) \cite{Raievi2011}. In addition to these bulk material observations of skyrmions, there have also been observations in ultra-thin films \cite{Heinze2011}. 

The envisaged technological application for which there has been hope and excitement is storing information using skyrmions in nanoribbons with a racetrack memory-like setup where the skyrmions can be moved using exceedingly small current densities \cite{Fert2013}. In addition to creating and moving these particle-like magnetic textures, there have been theoretical proposals for skyrmion logic gates \cite{Zhang2015} and skyrmion transistors \cite{Zhang2015-2}. Using metallic materials with regular, dissipative electrical currents leads to heating and less energy efficient devices. If, however, the low-dissipation currents in e.g. a topological insulator (TI) are used \cite{Hasan2010,RevModPhys.83.1057}, device efficiency can be increased. In addition to hosting electrical currents with its topologically protected edge states, the TI has spin-orbit coupling (SOC) that is transmitted to the magnetic material via the proximity effect.

Heterostructures of AFMs or ferromagnets (FMs) and topological insulator have been studied to some extent. Some works have investigated how the TI is affected by the proximity of an AFM \cite{Erlandsen2020}, whereas others have studied FM + TI and predicted induced DMI in the FM \waschanged{\cite{Nogueira2018,Scholten2021}} \waschanged{and electric-field-induced anisotropy fields affecting the spin dynamics \cite{Garate2010}}. The studies on the FM + TI system were conducted using a continuum model, in one case with temperatures above the Curie temperature of the FM \cite{Nogueira2018}. Self-energy effects on the surface states of the TI originating with spin-fluctuations on the FM have also been considered, within a lattice model \cite{Maeland2021}. Subsequently, the van der Waals Magnet \ch{Fe3GeTe2} epitaxially grown on a three-dimensional TI has been studied experimentally, displaying room-temperature Curie temperatures for monolayer FM \cite{Wang2020}. Another theoretical study focused on the magnetization dynamics at zero chemical potential, i.e., inside the magnetization-induced gap of the Dirac fermions \cite{Rex2016}.

In this paper, we study theoretically a heterostructure of FM+TI and ask to what extent spin interactions that promote non-collinear spin textures, including skyrmions, are generated by the proximity to a TI. In this way, one may hope to have a design of a platform  for generating ensembles of skyrmions, including skyrmion liquids or lattices \cite{Gruber2025}, using magnetic materials that are ubiquitous.       
We extend previous works to study the effects of the \emph{gapped} Dirac fermions in the TI on the FM, i.e., at temperatures below the Curie temperature, including also a variable chemical potential, which could tune the surface states of the TI in and out of the gapped state. This then also in principle allows tuning of the induced spin interactions {\it in situ}. Furthermore, we employ field theory techniques and approximations relevant in condensed matter systems. In particular, we do \emph{not} assume a Lorentz-invariant system in order to apply dimensional regularization, but rather we use a continuum theory with a short-distance cutoff that is of the order of the lattice constant, $a$. Furthermore, this allows us to treat the systems at finite temperature. We show that particle-hole excitations in the TI can mediate spin-spin interactions of various forms, such as exchange interactions and, importantly for skyrmion applications, DM interactions.

In the simple case of temporally local two-spin interactions, we find an RKKY-like spatial pattern \cite{Ruderman1954,Kasuya1956,Yosida1957}, where the scale of oscillations is set by the Fermi momentum, which can be tuned by changing the chemical potential through e.g. an electric field or chemical doping. The interaction strengths can vary considerably depending on the TI's Fermi velocity and the interfacial exchange coupling, ranging from the sub-$\mathrm{meV}$ range to around $\SI{1}{\milli \electronvolt}$. \waschanged{In addition, there is a sharp switching effect from finite induced RKKY interactions to zero whenever the chemical potential is inside the electronic gap induced by the magnetization. This generalizes a previous study \cite{Zyuzin2014} to intermediate chemical potentials and to inducing a gap by using a ferromagnetic platform instead of a superconducing one, the latter of which requires considerably lower temperatures.} 

Studying the dynamics of the FM, we find a nutation term \cite{Ciornei2011} in the magnetization equation of motion similar to that induced by normal-metal conduction electrons \cite{Kikuchi2015,Johnsen2025}, but its sign changes when the chemical potential passes through the gap. \waschanged{This causes the corresponding inertial spin-wave mode to be softened. }
Various other terms containing time and space derivatives and combinations thereof are also induced, but only a few are present to linear order in spin fluctuations \waschanged{around the uniform state}, where analytic expressions can be obtained. \waschanged{For nonuniform magnetization textures, there is a new induced term related to the time derivative of the magnetization's curl. This is relevant for skyrmion dynamics, such as their breathing mode, in which the skyrmion radius changes over time, thus changing the curl over time.}

This paper is organized as follows. In Section \ref{sec:Model}, we introduce the model we use. In Section \ref{sec:Interactions}, we give expressions for the induced spin-spin interactions in the FM due to the proximity to the TI, with details relegated to Appendix \ref{sec:path_integrals}. In Section \ref{sec:static_theory}, we focus on the static case and explicitly evaluate the radial and angular dependence of the antisymmetric contribution to the exchange tensor, along with the angular dependence of the diagonal and symmetric off-diagonal parts of the exchange tensor. The radial dependence of the latter two are treated in Appendix \ref{sec:static_appendix}. In Section \ref{sec:Taylor_expansion},  we perform an expansion to second order in frequency and momenta of the non-local in space and time generated coupling constants. In Section \ref{sec:Dynamics}, we use this expansion to obtain the equations of motion for the magnetization, as well as the renormalized magnon spectrum of the FM.   

\section{Model}
\label{sec:Model}
The model Hamiltonian is the sum
\begin{align}
    H = H_{\text{TI}} + H_\text{int} + H_\text{FM},
\end{align}
where we include an effective Hamiltonian for the TI's spin-momentum locked surface states, 
\begin{align}
    H_\text{TI} &=\int\dd[2]{r} c^\dagger(\rr) \left [ \tilde{\vb{d}}(-i\hbar \boldsymbol{\grad}) \cdot \bsigma - \mu \right] c(\rr)
    \label{eq:H_TI}
\end{align}
where the operator $\tilde{\vb{d}}$, which essentially stems from spin-orbit coupling, serves to model a zero-momentum Dirac cone and has the property $\tilde{\vb{d}}(-i\hbar \grad)^2 = -(\hbar \vF)^2 \grad^2$ \cite{Scholten2021}. The vector $\bsigma=(\sigma_x,\sigma_y,\sigma_z)$ consists of Pauli matrices, and $c^{(\dagger)}(\rr) = (c_\uparrow(\rr), c_\downarrow(\rr))^{(\dagger)}$ is a vector of annihilation (creation) operators $c^{(\dagger)}_\sigma(\rr)$ for an electron with spin $\sigma=\uparrow,\downarrow$ at position $\rr$. We consider an $sd$-like interfacial exchange coupling $\bar{J}>0$ between the Dirac fermions and the localized spins, 
\begin{align}\label{eq:H_int}
    H_\text{int} = -2\bar{J} \int \dd[2]{r}c^\dagger(\rr) \boldsymbol{\sigma} c(\rr) \cdot \bigl [ \vb{n}(\rr) + \tilde{m}_0\hat{\vb{z}} \bigr].
\end{align}
Here $\nn$ is a small fluctuation around the mean-field spin value $\tilde{m}_0 \hat{\vb{z}}$. The ferromagnet is thus assumed to be in a nearly spin-ordered state, meaning the Dirac fermions will be gapped. This spin order is stabilized by easy-axis anisotropy $K>0$ entering in the FM Hamiltonian,
\begin{align}\label{eq:FM_hamiltonian}
    H_\text{FM} = -\int \dd[2]{r} \left \{ K [\tilde{m}_0 + n_z(\rr)]^2 - J_\text{ex} [\boldsymbol{\grad} \nn(\rr)]^2 \right\},
\end{align}
where $J_\text{ex} > 0$ is the FM's intrinsic exchange interaction here taken to be spin-space isotropic. Writing out the above and introducing $m_0 = 2\bar{J}\tilde{m}_0$, we find
\begin{align}
    H_\text{TI} + H_\text{int} &= \int \dd[2]{r} c^\dagger(\rr) \left[ \vb{d}(-i\hbar \vF \grad)\cdot \boldsymbol{\sigma} -\mu \vphantom{\bar{J}} \right. \\\nonumber
    &\left.- 2\bar{J}\vb{n}(\rr) \cdot \boldsymbol{\sigma} \right] c(\rr).
\end{align}
Here, we combined $\tilde{\vb{d}}$ and $m_0$ into one operator $\vb{d}$. We study two specific forms of spin-orbit coupling in the TI, \cite{Nogueira2018}
\begin{align}
    \vb{d} &= \vb{d}_1 = \begin{pmatrix}
        -i\hbar \vF \partial_y\\
        i\hbar \vF \partial_x\\
        -m_0
    \end{pmatrix}
\label{eq:d1}    
\end{align}
and
\begin{align}
    \vb{d} &= \vb{d}_2 =  \begin{pmatrix}
        -i\hbar \vF \partial_x\\
        -i\hbar \vF \partial_y\\
        -m_0
    \end{pmatrix}.
    \label{eq:d2}
\end{align}
The labels are interchanged with respect to those of Ref.\ \cite{Nogueira2018}, where it was shown that $\vb{d}_2$ can give rise to Bloch skyrmions and $\vb{d}_1$ can give rise to Néel skyrmions.

We study the effects of the TI on the FM by integrating out the TI fermions in the path integral formalism. The computation is shown in Appendix \ref{sec:path_integrals}. We emphasize that we include the FM magnetization explicitly in Eq.\ \eqref{eq:H_int}, meaning we integrate out \emph{gapped} fermions $\psi, \psi^\dagger$ from the partition function
\begin{align}\label{eq:partition_function}
    Z = \int \mathcal{D}\vb{n}\e{-S_\text{FM}[\vb{n}]} \int \mathcal{D}\psi \mathcal{D}\psi^\dagger \e{-\psi^\dagger \left (G^{-1} + B\right)\psi}
\end{align}
where the fermion propagator $G$ is given by
\begin{align}
    G^{-1}(\rr, \partial_\tau) &= \partial_\tau - \mu + \vb{d}\cdot \bsigma,
\end{align}
and $B = -2\bar{J}\bsigma \cdot \vb{n}(\rr, \tau)$. In the Appendix, we find that the effective imaginary-time action is given by 
\begin{align}\label{eq:action_start}
    S_\text{eff}[\nn] &\approx S_\text{FM}[\nn] - \Tr \ln G^{-1} +\Delta S_\text{eff}^{(1)}[\nn] + \Delta S_\text{eff}^{(2)}[\nn],
\end{align}
where the third term,
$\Delta S_\text{eff}^{(1)}[\nn] = - \Tr (GB),$
is the first-order effect of magnetization ``reflected'' from the TI back into the FM, which looks like a homogenous field. It can therefore be subsumed into the magnetization, and we will not consider it further. The last term, 
\begin{align}\label{eq:def_s_eff_2}
    \Delta S_\text{eff}^{(2)}[\nn] = \frac{1}{2} \Tr (GBGB),
\end{align}
is our object of study, containing the two-spin interactions mediated by fermionic particle-hole excitations in the TI. The poles of $G$ give the dispersion relation
\begin{align}
    d_\pm(\kk) = -\mu \pm \sqrt{(\hbar \vF \kk)^2 + m_0^2}
\end{align}
for the TI fermions we integrate out, where the branches are separated by a gap $2m_0.$

\section{Induced spin-spin interactions}
\label{sec:Interactions}
The trace over the fermions in the generated two-spin action in Eq.\ \eqref{eq:def_s_eff_2} involves the trace over various products of Pauli matrices. In Appendix \ref{sec:path_integrals}, we show that the action can be simplified to
\begin{align}
    \Delta S_\text{eff}^{(2)}[\nn] &= \frac{1}{\beta} \sum_{i\omega_\nu} \int \frac{\dd[2]{q}}{(2\pi)^2} \\\nonumber
    &\times n_\alpha(\bq, i\omega_\nu) n_\beta(-\bq, -i\omega_\nu)\chi_{\alpha \beta}(\bq, i \omega_\nu),
\end{align}
where Matsubara frequencies with subscript $\nu$ are bosonic, i.e., $i\omega_\nu = 2\pi i \nu / \beta$ with the inverse temperature $\beta = 1/k_\text{B} T$ and integer $\nu$. This action has the rich spin structure
\begin{align}\nonumber
    &\chi_{\alpha \beta}(\bq, i\omega_\nu) = \frac{1}{\beta}\sum_n \int\frac{\dd[2]{k}}{(2\pi)^2}\\\nonumber
    &\times\Bigl[\delta_{\alpha \beta} J_\alpha(\kk+\bq, \kk, i\omega_n + i\omega_\nu, i\omega_n)  \\\label{eq:def_chi_alphabeta_q_omegaNu_structure}
    &+ (1-\delta_{\alpha\beta}) T_{\alpha \beta}^{\text{sym}}(\kk+\bq, \kk, i\omega_n + i\omega_\nu, i\omega_n) \\\nonumber
    &+ \epsilon_{\alpha \beta \gamma} D_\gamma(\kk+\bq, \kk, i\omega_n + i\omega_\nu, i\omega_n)\Bigr]
\end{align}
and is the object from which all results in this paper are derived. The Matsubara frequencies with Latin subscripts, $i\omega_n = \pi i(2n+1)/\beta$ for integer $n$, are fermionic. The interaction coefficients in their general form are given by
\begin{widetext}
\begin{subequations}
\label{eq:DTJ_general_start}
\begin{align}
    D_\gamma(\kk_1, \kk_2, i\omega_{n_1}, i\omega_{n_2}) &= 4i\bar{J}^2 \frac{(i\omega_{n_2} + \mu)d_\gamma(\kk_1) - (i\omega_{n_1} + \mu)d_\gamma(\kk_2)}{\left [(i\omega_{n_1} + \mu)^2 - d^2(\kk_1) \right]\left [(i\omega_{n_2} + \mu)^2 - d^2(\kk_2) \right]} \\ \label{eq:T_momenta_freqs_start}
    T^\text{sym}_{\alpha \beta}(\kk_1, \kk_2, i\omega_{n_1}, i\omega_{n_2}) &= 4\bar{J}^2 \frac{d_\alpha(\kk_1) d_\beta(\kk_2) + d_\beta(\kk_1) d_\alpha(\kk_2)}{\left [(i\omega_{n_1} + \mu)^2 - d^2(\kk_1) \right]\left [(i\omega_{n_2} + \mu)^2 - d^2(\kk_2) \right]}\\ \label{eq:J_momenta_freqs_start}
    J_\alpha(\kk_1, \kk_2, i\omega_{n_1}, i\omega_{n_2}) &= 4\bar{J}^2 \frac{(i\omega_{n_1} + \mu) (i \omega_{n_2} + \mu) + d_\alpha(\kk_1) d_\alpha(\kk_2) - \sum_{\gamma \neq \alpha} d_\gamma(\kk_1) d_\gamma(\kk_2)}{\left [(i\omega_{n_1} + \mu)^2 - d^2(\kk_1) \right]\left [(i\omega_{n_2} + \mu)^2 - d^2(\kk_2) \right]}.
\end{align}
\end{subequations}
\end{widetext}
The $J_\alpha$ term enters as an anisotropic Heisenberg exchange interaction, where the anisotropies stem from both SOC in the TI and the magnetization in the FM. Setting the Dirac fermions in the TI apart from those in a normal metal \cite{Johnsen2025} is their off-diagonal spin-space structure, which is transmitted to the FM through the interfacial exchange coupling $\bar{J}$. The first off-diagonal term, $D_\gamma$, resembles a DM interaction in its antisymmetric spin structure, and we will show that it indeed takes this form in the long-wavelength limit. The off-diagonal symmetric $T^\text{sym}_{\alpha \beta}$ term is also present and was shown to affect the stability of Néel skyrmions in TI-FM heterostructures in the long-wavelength limit \cite{Nogueira2018}.

One of the motivations of this paper is that, so far, this calculation does not assume $\beta \to \infty$ nor does it try to keep Lorentz invariance and Poincaré symmetry to use dimensional regularization to compute the $\kk$ integral in $\chi_{\alpha \beta}(\bq, i\omega_\nu)$. Such symmetries are not present in condensed matter systems, where instead there is an underlying lattice and thus a short-distance cutoff, $a$, the lattice constant. The treatment presented here should be the correct one in a condensed-matter setting, and we will highlight the additional structure appearing in the equation of motion for the magnetization $\nn$ when taking the underlying lattice into account.

This paper contains two distinct calculations that deal with the induced interactions $J_\alpha, D_\gamma, T_{\alpha \beta}^\text{sym}$, so we outline the difference between the two here. One is a common procedure \cite{Rex2016,Nogueira2018} used to make a simple equation of motion for the magnetization. The starting point is a Taylor expansion of
\begin{align}\label{eq:taylor_imaginary}
    &\chi_{\alpha\beta}(\bq, i\omega_\nu) = \chi_{\alpha \beta}^{(0,0)} + \chi_{\alpha \beta \gamma}^{(0,1)} q_\gamma + \chi_{\alpha \beta}^{(1,0)} i\omega_\nu\\\nonumber
    & + \chi_{\alpha \beta \gamma}^{(1,1)}i\omega_\nu q_\gamma + \chi_{\alpha \beta \gamma \delta}^{(0,2)} q_\gamma q_\delta+ \chi_{\alpha \beta}^{(2,0)} (i\omega_\nu)^2 + \dots
\end{align}
around $\bq = 0, i\omega_\nu = 0$. By truncating this series at some order -- we will go to second order -- one extracts the model's behavior in the long-wavelength--near-static limit, yielding a spatially and temporally local equation of motion. It will allow us to study the FM's dynamics and spin-wave dispersion in Section \ref{sec:Dynamics}.

The other calculation pertains to the RKKY-like \emph{spatial} dependence of the induced spin-spin interactions in the \emph{static} case $\chi_{\alpha \beta}(\bq, i\omega_\nu=0)$, which in the notation of Eq.\ \eqref{eq:taylor_imaginary} means keeping only the terms $\chi^{(0,n)}$ for all integers $n\geq 0$. Importantly, we treat the function $\chi_{\alpha \beta}(\bq, i\omega_\nu=0)$ explicitly in Section \ref{sec:static_theory}, not its Taylor-expanded form. The latter would lead to an infinite-order spatial gradient expansion. Instead, we keep the Fourier transformed action 
\begin{align}\nonumber
    &\Delta S_\text{eff}^{(2)}[\nn] = \frac{1}{\beta} \sum_{i\omega_\nu} \int \dd[2]{r_1} \dd[2]{r_2}  n_\alpha(\rr_1, i\omega_\nu) n_\beta(\rr_2, -i\omega_\nu) \\\label{eq:action_matsubara_r_general}
    &\times\chi_{\alpha \beta}(\rr_2 - \rr_1, i\omega_\nu)
\end{align}
and consider the $i\omega_\nu=0$ term. The above is an action expressed in terms of Matsubara frequencies instead of real frequencies. To obtain the real-frequency static interaction, one needs only replace $\beta \to 2\pi$. This is due to the fact that, for temperatures low enough to replace $\frac{1}{\beta} \sum_{i\omega_\nu} \to \int \frac{\dd{\omega}}{2\pi}$, the real-time action is obtained from
\begin{align}\label{eq:imag_to_real_time_conversion_action}
    \e{i\Delta S_\text{eff}^{\text{(2, RT)}}} &= \left. \e{-\Delta S_\text{eff}^{(2)}} \right |_{i \omega \to \omega + i \delta},
\end{align}
where $\delta$ is some infinitesimal quantity to avoid integrating over mathematical poles. In general, the result is
\begin{align}
    &\Delta S_\text{eff}^{\text{(2, RT)}} = \int \frac{\dd{\omega}}{2\pi} \frac{\dd[2]{q}}{(2\pi)^2}n_\alpha(\bq, \omega ) \biggl[\chi_{\alpha \beta}^{(0)}(\bq) \\\nonumber
    &+ \omega \chi_{\alpha \beta}^{(1)}(\bq) + \dots \biggr] n_\beta(-\bq, -\omega ),
\end{align}
where we temporarily introduced the $\omega$ Taylor coefficients $\chi_{\alpha \beta}^{(n)}(\bq)$. We will only consider $n=0$ in Section \ref{sec:static_theory}, so the corresponding static action in real space has the form
\begin{align}\label{eq:static_action_realSpace_general}
    &\Delta S_\text{eff}^{\text{(2, RT, static)}} =\int\dd{t}\int \dd[2]{r_1} \dd[2]{r_2} n_\alpha(\rr_1, t) n_\beta(\rr_2, t)\\\nonumber
    &\times \frac{\chi_{\alpha \beta}(\rr_2 - \rr_1, i\omega_\nu=0)}{2\pi},
\end{align}
where we Fourier transformed to real time $t$ using a symmetric-prefactor convention. Here, the quantity $\chi_{\alpha \beta}(\rr_2  - \rr_1, i\omega_\nu=0)$ is the same one as in Eq.\ \eqref{eq:action_matsubara_r_general}.

In contrast to the above, in Sections \ref{sec:Taylor_expansion} and \ref{sec:Dynamics} we will use the gradient and time-derivative expansion picture. In that view, one replaces all the $\omega$'s and $q_\gamma$'s by derivatives in the expansion of $\chi_{\alpha \beta}(\bq, i\omega_\nu)$ shown in Eq.\ \eqref{eq:taylor_imaginary},
\begin{widetext}
\begin{align}\label{eq:action_realtime_rt}
     \Delta S_\text{eff}^{(\text{2, RT})} &= \int\dd[2]{r_1} \dd[2]{r_2} \int\frac{\dd t_1 \dd t_2}{2\pi} \int\frac{\dd \omega}{2\pi} \int \frac{\dd[2]{q}}{(2\pi)^2} n_\alpha(\rr_1, t_1) \e{-i(\bq\cdot \rr_1 - \omega t_1)} \biggl [\chi_{\alpha\beta} \left(\frac{1}{i} \boldsymbol{\grad}_{\rr_2}, \frac{1}{-i} \partial_{t_2}\right)  \e{i(\bq \cdot \rr_2 - \omega t_2)} \biggr ] n_\beta(\rr_2, t_2).
\end{align}
\end{widetext}
Each term in the expansion must be treated separately, integrating by parts the appropriate number of times to move the derivatives from the exponential to $n_\beta$. This results in a spatially and temporally local action \footnote{apart from the derivatives, which are technically not local.}, i.e., an action containing only one space and one time coordinate.

\section{Static Theory}
\label{sec:static_theory}
This section contains the results of the static limit of our model, which is an extension of Ref.s \cite{Nogueira2018,Rex2016} to nonlocal spin-spin interactions. They can be viewed as generalized RKKY interactions between spins separated by a displacement vector $\Delta \rr$ mediated by the topological electrons in the TI. We calculate the static susceptibility, or Lindhard function, from Eq.\ \eqref{eq:def_chi_alphabeta_q_omegaNu_structure} in real space, $\chi_{\alpha \beta}(\Delta \rr, i\omega_\nu=0)/2\pi$. The factor $2\pi$ is needed in the real-frequency context, as shown in Eq.\ \eqref{eq:static_action_realSpace_general}. As always, the structure of the susceptibility is
\begin{align}\nonumber
    \chi_{\alpha \beta}(\Delta \rr, i\omega_\nu) &= \delta_{\alpha \beta} J_\alpha(\Delta \rr, i\omega_\nu) + (1-\delta_{\alpha \beta}) T^{\text{sym}}_{\alpha \beta}(\Delta \rr, i\omega_\nu) \\ 
    &+ \epsilon_{\alpha \beta \gamma} D_\gamma (\Delta \rr, i\omega_\nu),
\end{align}
where each term consists of two momentum integrals and a Matsubara sum over the expressions in Eq.\ \eqref{eq:DTJ_general_start}. Fortunately, as can be seen from Eq.\ \eqref{eq:DTJ_general_start}, the integrand can be written as a sum of terms where each term separates in momenta. The difficult part turns out to be the Matsubara sum, where at zero temperature we are able to get an analytic expression only for the DMI-like RKKY interaction $D_\gamma$. The exchange and symmetric off-diagonal interactions $J_\alpha, T_{\alpha \beta}^{\text{sym}}$ are reduced to energy integrals that can be computed numerically. An overview of these lengthy calculations is given in Appendix \ref{sec:static_appendix}.

The analytical, zero-temperature expression for the DMI term is
    \begin{align}
        &D_\gamma(\Delta \rr, i\omega_\nu=0) = \waschanged{\sgn (\mu) \theta(\abs{\mu} - m_0)} \frac{\bar{J}^2 \abs{m_0}^3}{(\hbar \vF)^4}  \\\nonumber
        &\times\Phi^D_\gamma(\phi)\frac{\left(\frac{\mu^2}{m_0^2} -1\right)^{3/2}}{2\pi \abs{\Delta \bar{\rr}}} J_1\left(\abs{\Delta \bar{\rr}}\right) Y_1\left(\abs{\Delta \bar{\rr}} \right),
    \end{align}
where we defined the dimensionless coordinate
\begin{align}\label{eq:r_bar_def}
     \Delta \bar{\rr} \equiv \Delta \rr \frac{\sqrt{\mu^2 - m_0^2}}{\hbar \vF} =\Delta \rr \kF
\end{align}
and the polar angle $\phi$ between the $x$ axis and $\Delta \rr$, showing that the length scale of the oscillations is set by the Fermi momentum. This fact becomes more transparent in the long-range limit $\abs{\Delta \bar{\rr}} \gg 1$, 
\begin{align}
    D_\gamma(\abs{\Delta \rr} \gg 1/\kF, i\omega_\nu=0) \propto \Phi_\gamma^D(\phi) \frac{\cos (2\abs{\Delta \bar{\rr}})}{\Delta \bar{\rr}^2},
\end{align}
which displays the same $\abs{\Delta \rr}$ dependence as RKKY interactions mediated by a two-dimensional gas of electrons with parabolic dispersion \cite{Aristov1997,Litvinov1998}. The vector $\boldsymbol{\Phi}$ contains the angular dependence of $D_\gamma$ and is given by
\begin{align}
    \boldsymbol{\Phi}^{D,1}(\phi) = \begin{pmatrix}  -\sin \phi\\ \cos \phi \\0 \end{pmatrix}, \boldsymbol{\Phi}^{D,2}(\phi) = \begin{pmatrix}  -\cos \phi\\ -\sin \phi \\0 \end{pmatrix}
\end{align}
for the choice of SOC $\vb{d}_1$ and the choice $\vb{d}_2$, respectively. This nontrivial angular dependence is caused by the TI's SOC, and leads to $p$-wave spatial symmetry in the DMI term. \waschanged{The dominant parts, lying on the coordinate axes, are of the form
\begin{align}
    \bhat{x}&\cdot [\nn(\rr) \cross \nn(\rr + \Delta r \bhat{y})] \\\nonumber
    &- \bhat{y}\cdot[\nn(\rr) \cross \nn(\rr + \Delta r \bhat{x})]
\end{align}
for $\vb{d}_1$, favoring Néel skyrmions with no in-plane rotation, and of the form
\begin{align}
    \bhat{x}&\cdot [\nn(\rr) \cross \nn(\rr + \Delta r \bhat{x})] \\\nonumber
    &+\bhat{y}\cdot[\nn(\rr) \cross \nn(\rr + \Delta r \bhat{y})]
\end{align}
for $\vb{d}_2$, favoring Bloch skyrmions, which do have in-plane rotation \cite{Leonov2016}. This conclusion is the same as that of Ref.~\cite{Nogueira2018}, where it should be noted that the definitions of $\vb{d}_1, \vb{d}_2$ are switched compared to the definitions herein.}

\begin{figure*}
    \centering
    \begin{subfigure}[b]{0.32\textwidth}
        \centering
        \includegraphics[width=\textwidth]{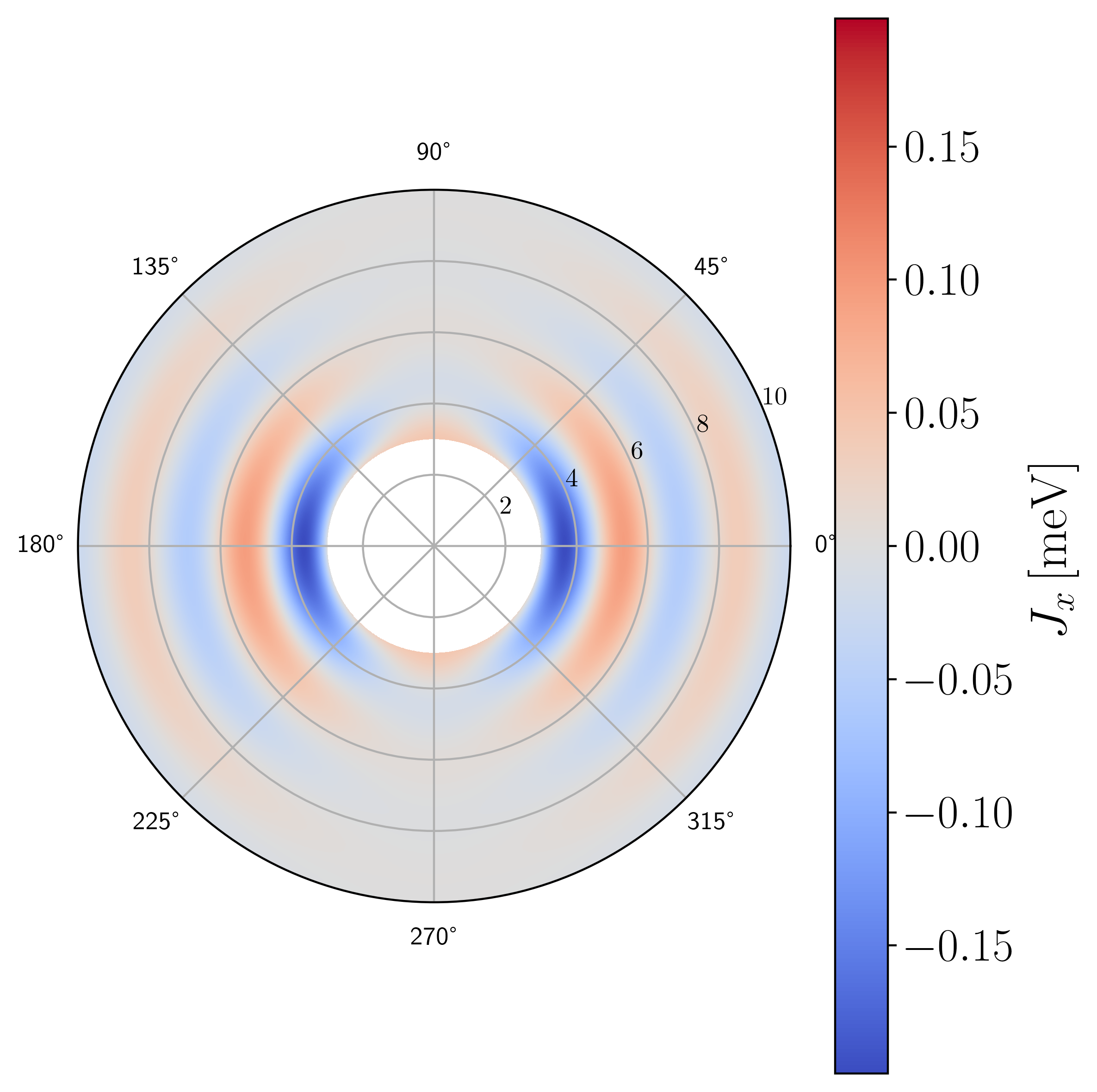}
    \end{subfigure}
    \centering
    \begin{subfigure}[b]{0.32\textwidth}
        \centering
        \includegraphics[width=\textwidth]{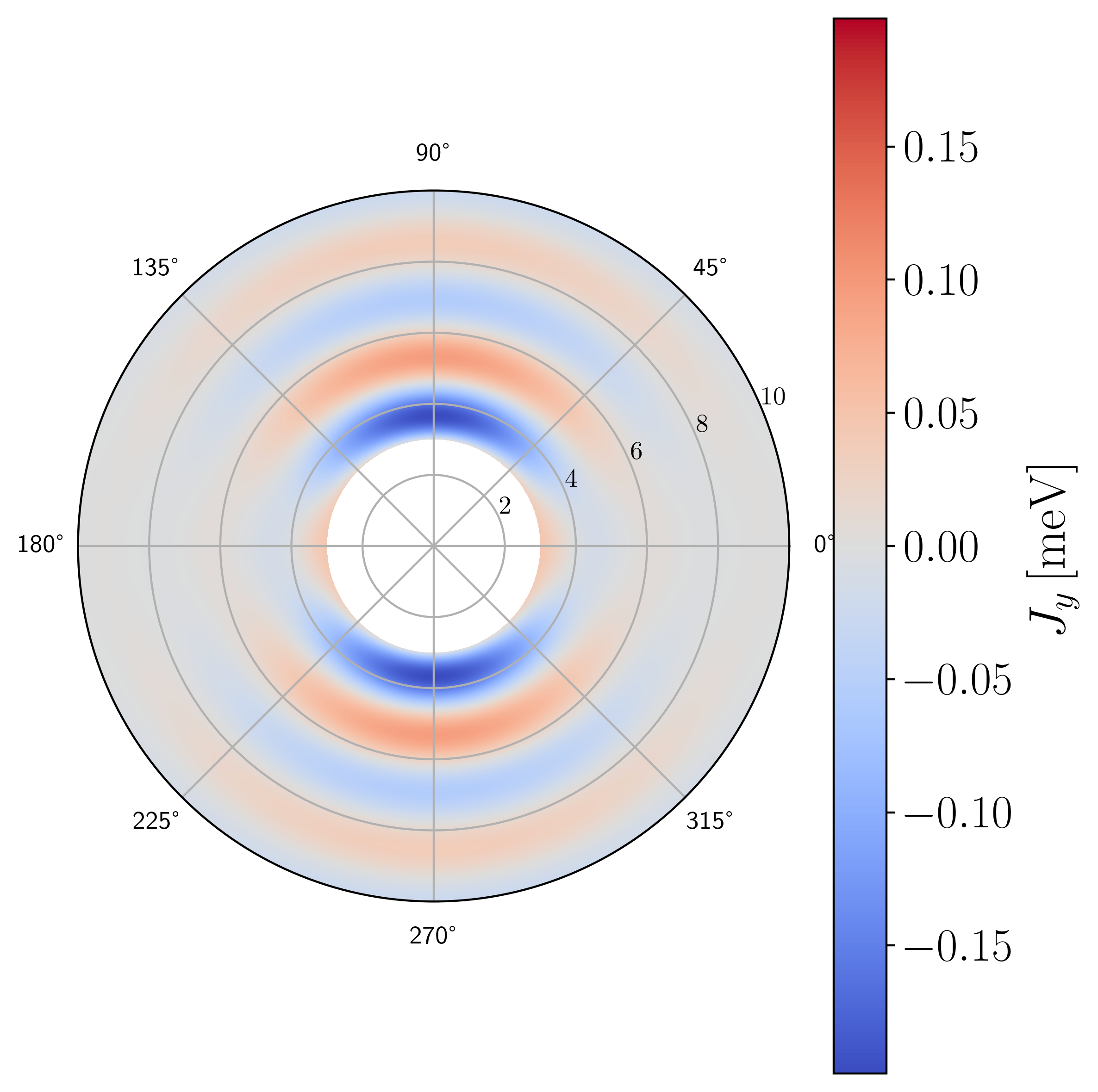}
    \end{subfigure}
    \centering
    \begin{subfigure}[b]{0.32\textwidth}
        \centering
        \includegraphics[width=\textwidth]{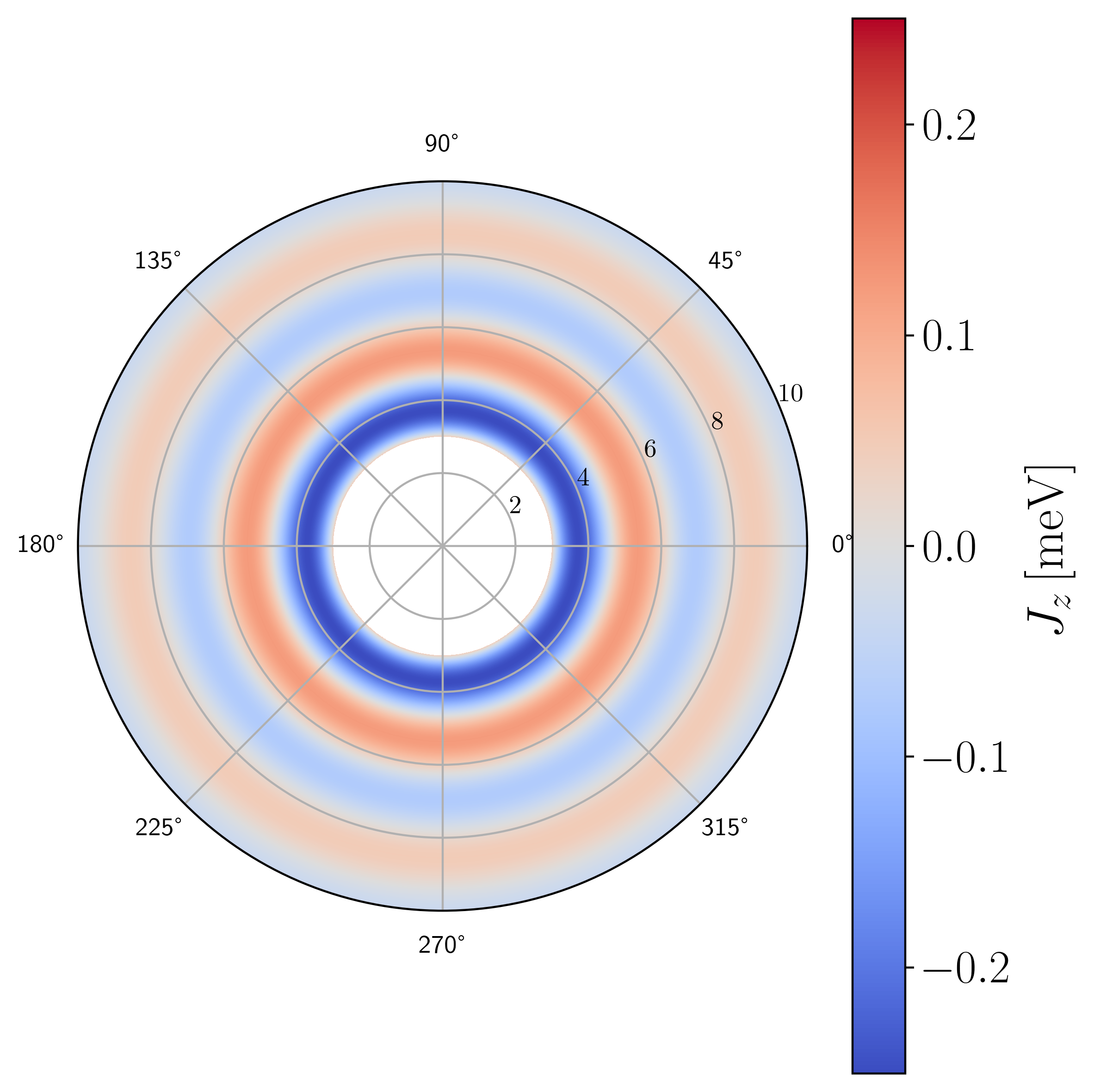}
    \end{subfigure}
     \centering
     \begin{subfigure}{0.32\linewidth}
        \centering
        \includegraphics[width=\linewidth]{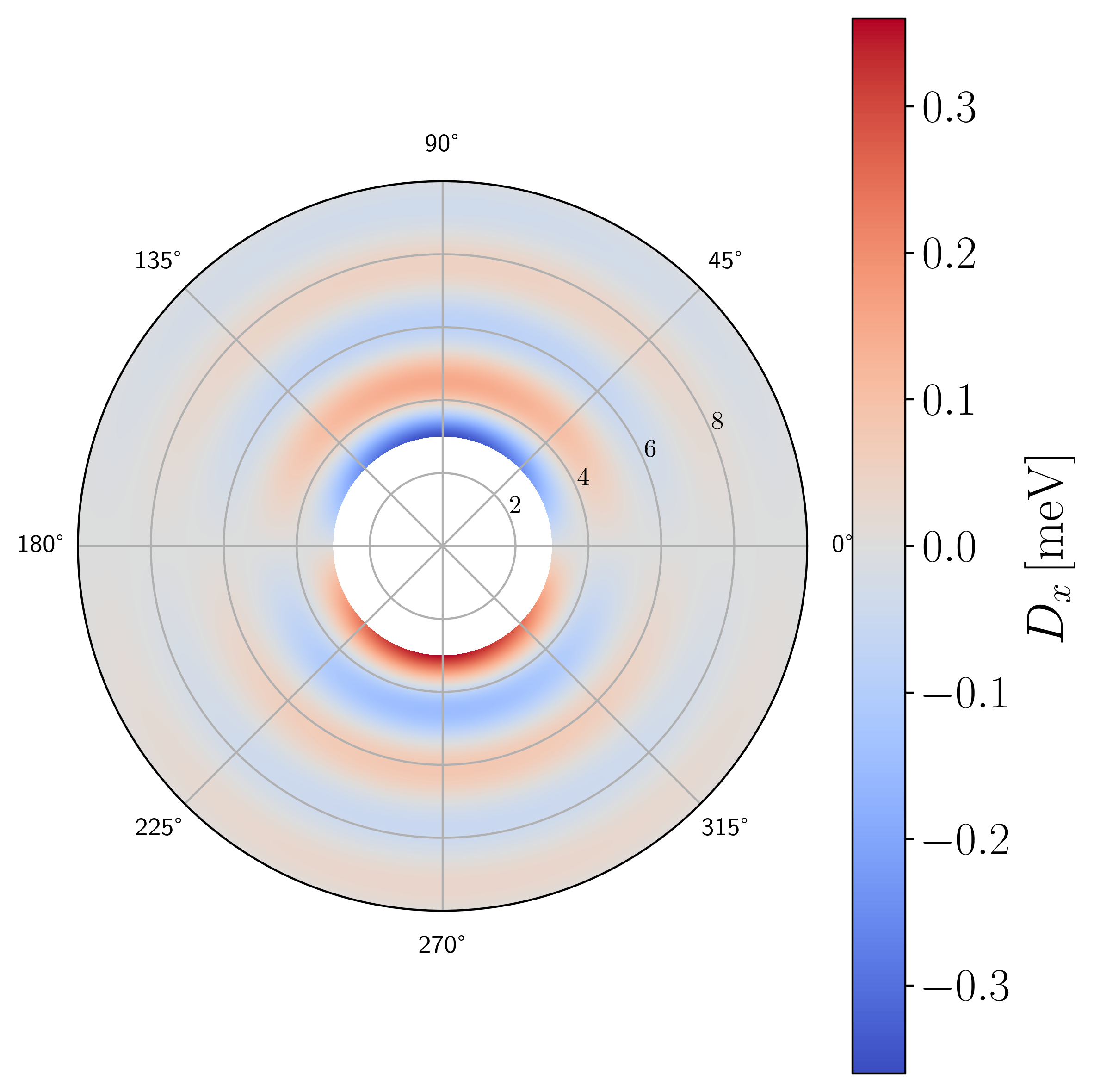}
    \end{subfigure}
     \centering
     \begin{subfigure}{0.32\linewidth}
        \centering
        \includegraphics[width=\linewidth]{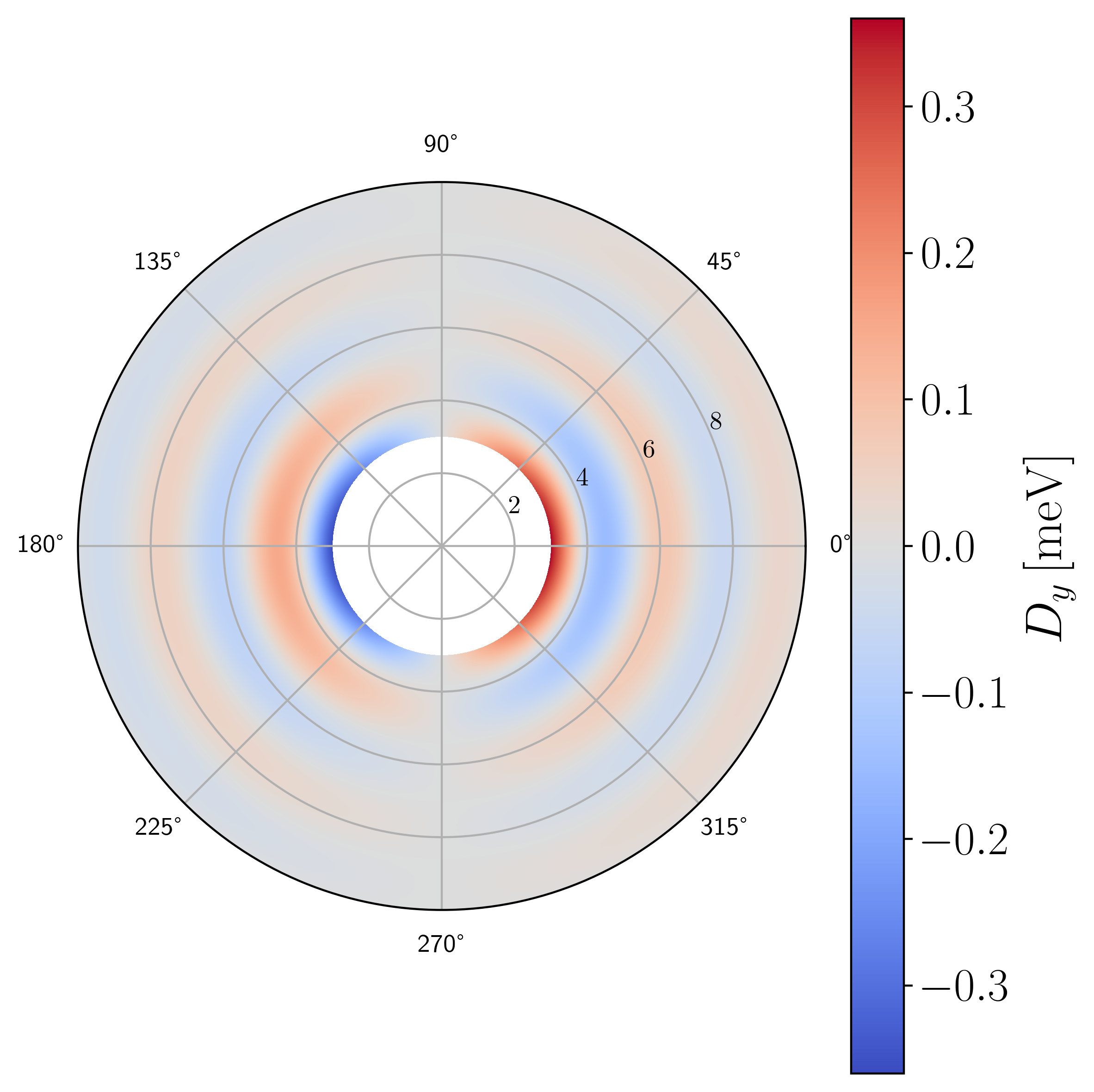}
    \end{subfigure}
     \centering
     \begin{subfigure}{0.32\linewidth}
        \centering
        \includegraphics[width=\linewidth]{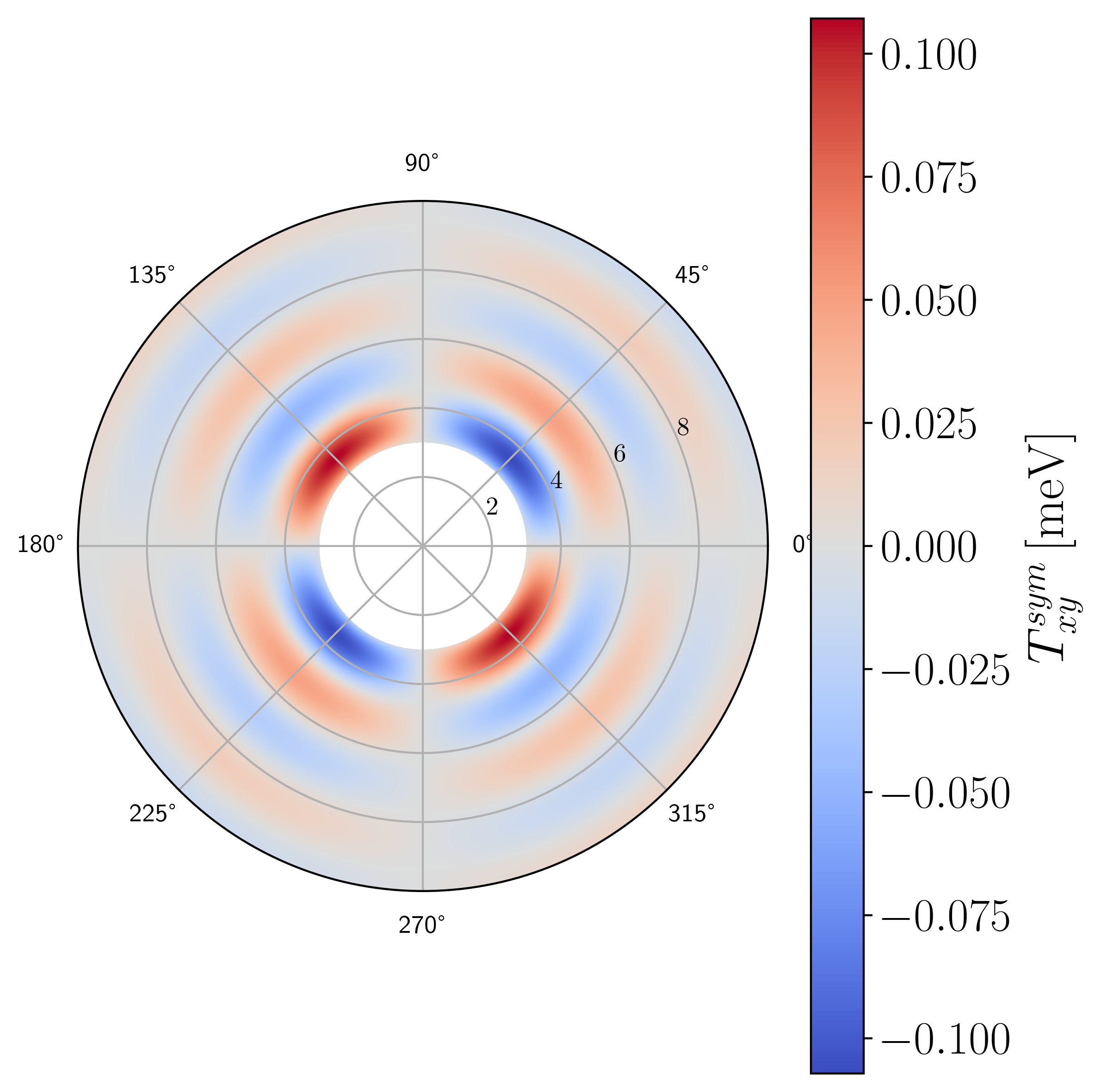}
    \end{subfigure}
    \caption{\small Summary of the static spin theory in real space $\Delta \rr$, containing induced anisotropic Heisenberg exchange interactions $J_\alpha \delta_{\alpha \beta}(\Delta \rr)$, Dzyaloshinskii–Moriya interactions $D_\gamma(\Delta \rr) \epsilon_{\gamma \alpha \beta}$, and an off-diagonal spin-symmetric interaction $T_{\alpha \beta}^\text{sym}(\Delta \rr)(1-\delta_{\alpha \beta}).$ The displacement vector $\Delta \rr$ connects two spin components, $S_\alpha(\rr)$ and $S_\beta(\rr + \Delta \rr)$, and is given in units of the lattice constant $a$. The numerical values shown are obtained at chemical potential $\mu=\SI{300}{\milli \electronvolt},$ the Fermi velocity given by $\hbar \vF/a=\SI{200}{\milli \electronvolt}$, and interfacial exchange coupling $\bar{J}=\SI{75}{\milli \electronvolt}$ using the choice of spin-orbit coupling (SOC) $\vb{d}_1$. The other choice $\vb{d}_2$ can be obtained by shifting the polar coordinate $\phi\to \phi+\pi/2$. The amplitudes remain the same for both kinds of SOC.}
    \label{fig:Jalpha_summary}
\end{figure*}

A plot of the spatial structure for $m_0=\SI{20}{\milli \electronvolt}$ and $\mu = \SI{200}{\milli \electronvolt}$ is shown in Fig.\ \ref{fig:Jalpha_summary}. It is noteworthy that $D_z(\Delta \rr, i\omega_\nu=0) = 0$, which is the component of the DM interaction that causes canting in-plane, i.e., the $n_x(\rr_1) n_y(\rr_2) - n_y(\rr_1) n_x(\rr_2)$ coefficient.  This property is not specific to the particular form of the spin-momentum locking term ${\vb{d}}(-i\hbar \boldsymbol{\grad}) \cdot \bsigma$ in Eq. \eqref{eq:H_TI}. Instead, it arises from the lack of dispersion in the $z$ direction, which implies that two-dimensional interfaces, such as the one considered here, do not admit an induced out-of-plane component of the DM vector $\vb{D}(\Delta \rr)$.

Through symmetry arguments, it can be shown that the static off-diagonal symmetric interaction has the property
\begin{align}
    T_{\alpha z}^\text{sym}(\Delta \rr, i\omega_\nu=0)=0,
\end{align}
whereas $T_{xy}^\text{sym}(\Delta \rr, i\omega_\nu = 0)$, the coefficient of $n_x(\rr_1) n_y(\rr_2) + n_y(\rr_1) n_x(\rr_2)$, is finite. The expression for the latter is shown in Appendix \ref{sec:static_appendix}, where it is simplified to one-dimensional integrals over Bessel functions. It has $d$-wave symmetry in the $\Delta \rr$ plane, 
\begin{align}
T_{xy}^\text{sym}(\Delta \rr, i\omega_\nu = 0) \propto \pm \sin (2 \phi),
\end{align}
where the upper sign corresponds to $\vb{d}_1$ and the lower sign to $\vb{d}_2$. \waschanged{It is even in $\mu$ and vanishes for $\mu$ in the gap.}

Also the induced exchange interaction terms $J_\alpha$ have too large expressions to be printed here, but all of them are finite. They have the structure
\begin{align}
    J_\alpha(\Delta \rr, i\omega_\nu = 0) \propto \mathcal{A}_\alpha(\abs{\Delta \rr}) + \mathcal{B}_\alpha(\abs{\Delta \rr}) \Phi_\alpha^{J}(\phi),
\end{align}
where $\mathcal{A}_\alpha, \mathcal{B}_\alpha$ are complicated expressions given in Appendix \ref{sec:static_appendix}\waschanged{, and their $\mu$ dependence is as for $T^\text{sym},$ i.e., they are proportional to $\theta(\abs{\mu}-m_0)$}. The angular dependence is given by
\begin{align}
    \boldsymbol{\Phi}^{J}(\phi) = \begin{pmatrix}
         \mp\cos 2\phi\\
        \pm \cos 2\phi\\
        -1
    \end{pmatrix},
\end{align}
where the upper sign corresponds to $\vb{d}_1$ and the lower sign to $\vb{d}_2$. Thus, the variation $\Phi_{\alpha\neq z}^J(\phi)$ of $J_{\alpha \neq z}$ has $d$-wave symmetry, whereas $J_z$ has $s$-wave symmetry.
Due to the prefactor $\vF^4 / \bar{J}^5$, the numerical values of the peaks in $\chi_{\alpha \beta}(\Delta \rr, i\omega_\nu=0)$ can be less than $\SI{1}{\milli \electronvolt}$ with small $\bar{J} = \SI{35}{\milli \electronvolt}$ and large Fermi velocity $\frac{\hbar \vF}{a} = \SI{500}{\milli \electronvolt}$, or they can reach values nearing an electronvolt if a material with lower Fermi velocity $\vF$ is used in conjunction with enhanced $\bar{J}$. Large bandwidth and chemical potential also aid in increasing the size of the induced interactions. Changing $\mu, \bar{J}$ also changes the length scale over which the interactions switch sign because the difference $\mu^2 - m_0^2$ in Eq.\ \eqref{eq:r_bar_def} changes.

These results show that the induced RKKY interactions -- including DMI, which is highly relevant if this kind of heterostructure is to be used as a platform for skyrmions -- have a spatial structure with switching signs as a function of spin-spin distance, and the size of the amplitudes can be tuned by tuning $\mu$, or they can be enhanced by using materials with smaller $\vF$. \waschanged{Furthermore, all the interactions can be switched off by placing the chemical potential inside the gap, $\abs{\mu} < m_0$. This extends a previous study of RKKY interactions on the surface of a TI to intermediate chemical potentials and to gapping the TI electrons using a magnetic system instead of a superconductor \cite{Zyuzin2014}. The superconducting platform will in general require much lower temperatures than the magnetic one proposed here.}

\section{Second-order expansion}
\label{sec:Taylor_expansion}
The other kind of calculation we will show is based on truncating the Taylor expansion in Eq.\ \eqref{eq:taylor_imaginary}, keeping only terms of second order and lower,
\begin{align}
    \chi_{\alpha \beta}(\bq, i\omega_\nu) \approx \sum_{i+j+k\leq 2} q_x^i q_y^j (i\omega_\nu)^k \chi_{\alpha \beta}^{(k, i, j)}.
\end{align}
It is a generalization of Ref.\ \cite{Nogueira2018} to finite magnetization and frequency and a generalization of Ref.\ \cite{Rex2016} to taking the underlying Poincaré-symmetry breaking lattice into account. We will proceed by using the Euler-Lagrange equations to construct an equation of motion for the magnetization and compute the spin-wave dispersion relation of the system.

The Taylor coefficients are 
\begin{align}
    \chi_{\alpha \beta}^{(k, i, j)} &= \lim_{\bq \to 0, i\omega_\nu \to 0} \frac{1}{i! (j!) (k!)} \left(\partial_{q_x}\right)^i \left(\partial_{q_y}\right)^j \\\nonumber
    &\times                      \left(\partial_{(i\omega_\nu)}\right)^k \chi_{\alpha \beta}(\bq, i\omega_\nu)
\end{align}
by definition. The overview of the calculation is that we will bring the differential operators inside the $n$ sum and momentum integral,
\begin{align}\label{eq:taylor_coeff_order}
    &\chi_{\alpha \beta}^{(k, i, j)} =
    \frac{1}{\beta (i!) (j!) (k!)}\int\frac{\dd[2]{k}}{(2\pi)^2}  \sum_n \lim_{\bq \to 0, i\omega_\nu \to 0}  \\\nonumber
    &\left(\partial_{q_x}\right)^i \left(\partial_{q_y}\right)^j \left(\partial_{(i\omega_\nu)}\right)^k \chi_{\alpha \beta}(\kk+\bq, \kk, i\omega_n + i\omega_\nu, i\omega_n),
\end{align}
then compute the sum over $n$ for each type of interaction $J_\alpha,D_\gamma, T_{\alpha \beta}^\text{sym}$ using the Residue theorem. There are a few subtleties related to derivatives of the Fermi-Dirac distribution that are outlined along with some details of the computation of the $n$ sum and the $\kk$ integral in Appendix \ref{sec:dynamic_appendix}. The general structure of the induced exchange interactions is
\begin{align}\label{eq:J_alpha_expansion}
    J_\alpha(\bq, i\omega_\nu) &= J^{(0,0)}_{\alpha} + J^{(2,0)}_{\alpha} (i\omega_\nu)^2 + J^{(1,0)}_{\alpha} i\omega_\nu \\\nonumber
    &+ J^{(0,2)}_{\alpha, xx} q_x^2 + J^{(0,2)}_{\alpha, yy} q_y^2.
\end{align}
The off-diagonal symmetric interaction has different kinds of terms for each component, namely
\begin{subequations}
\begin{align}
    T_{xy}^\text{sym}(\bq, i\omega_\nu) &= T_{xy, xy}^{(0,2)} q_x q_y \\
    T_{xz}^\text{sym}(\bq, i\omega_\nu) &= T_{xz, y}^{(1,1)} i\omega_\nu q_y + T_{xz, y}^{(0,1)} q_y,
\end{align}
\end{subequations}
and $T_{yz}^\text{sym}(\bq, i\omega_\nu)$ can be obtained from $T_{xz}^\text{sym}(\bq, i\omega_\nu)$ by replacing $q_y \rightarrow - q_x$. \waschanged{To our knowledge, the $T^{(1,1)}$ term has not been reported previously for this system. It is given an interpretation in next section.} Finally, the DMI terms are given by
\begin{subequations}\label{eq:D_gamma_expansion}
\begin{align}
    D_x(\bq, i\omega_\nu) &= D_{x, y}^{(1,1)} i\omega_\nu q_y + D_{x, y}^{(0,1)} q_y\\
    D_z(\bq, i\omega_\nu) &= D_{z}^{(2,0)} (i\omega_\nu)^2 + D_{z}^{(1,0)} i\omega_\nu,
\end{align}
\end{subequations}
and $D_y(\bq, i\omega_\nu)$ can be obtained from $D_x(\bq,i\omega_\nu)$ by replacing $q_y \rightarrow - q_x$. All the coefficients depend on model parameters such as the chemical potential $\mu$, the Fermi velocity $\vF$, the magnetization $m_0$, and the interfacial exchange interaction strength $\bar{J}$. They take on constant values as functions of $\mu$ when $\abs{\mu}<m_0$, i.e., when the chemical potential is inside the gap. In fact, $J_\alpha^{(1,0)}$, $T_{\alpha z}^{(0,1)}$, \waschanged{$T^{(1,1)}_{\alpha z}$,} $D_{\alpha \neq z}(\bq, i\omega_\nu)$ to second order, and $D_{z}^{(2,0)}$ all vanish inside the gap, explaining how these new kinds of terms in the Lagrangian appear here, but not in Ref.\ \cite*{Rex2016}, where $\mu = 0$. The full form of the coefficients is shown in Appendix \ref{sec:dynamic_appendix}.

Our results show anisotropic induced exchange interactions. In particular, due to breaking Lorentz symmetry and time-reversal symmetry, the Lagrangian density does not contain 
\begin{align}
     &s\left [ (\grad \cdot \nn)^2 + (\grad \nn)^2 \right] \\\nonumber
     & = s \left\{ 2\left[ (\partial_x n_x)^2 + (\partial_y n_y)^2 + (\partial_x n_x)(\partial_y n_y) \right] + (\partial_y n_x)^2 \right. \\\nonumber
     &\left.+ (\partial_x n_y)^2 + (\partial_x n_z)^2 + (\partial_y n_z)^2 \right \}
\end{align}
with the same prefactor $s$ in front of all the derivative terms, but rather 
\begin{align}\nonumber
    &T_{xy,xy}^{(0,2)} 2 (\partial_x n_x) (\partial_y n_y)+J_{x,xx}^{(0,2)} \left( (\partial_x n_x)^2 + (\partial_y n_y)^2 \right)\\
    & + J_{x,yy}^{(0,2)} \left( (\partial_y n_x)^2 +  (\partial_x n_y)^2 \right) \\\nonumber
    &+ J_{z,xx}^{(0,2)} \left ( (\partial_x n_z)^2 + (\partial_y n_z)^2 \right),
\end{align}
where the prefactors are distinct for $\mu$ outside the gap.

Next we study the consequences these additional terms have for the dispersion of linear spin waves.

\section{Equation of motion}
\label{sec:Dynamics}
Referring to Appendix \ref{sec:llg_appendix}, the generated two-spin part of the real-time action is
\begin{align}
     &\Delta S_\text{eff}^{(\text{2, RT})} = \int\dd[2]{\rr} \int \dd t \mathcal{L}^{(2)}_\text{eff}
\end{align}
where the Lagrangian density has the form
\begin{align}\label{eq:lagrangian_for_eom}
    \mathcal{L}^{(2)}_\text{eff} &= \frac{1}{2\pi}n_\alpha(\rr, t) \biggl [ \chi_{\alpha \beta}^{(0,0)} + i \chi_{\alpha \beta \gamma}^{(0,1)} \partial_{\gamma}- i \chi_{\alpha \beta }^{(1,0)}  \partial_{t} \\\nonumber
     & + \chi_{\alpha \beta \gamma}^{(1,1)}\partial_{t} \partial_{\gamma} - \chi_{\alpha \beta \gamma \delta}^{(0,2)} \partial_{\gamma} \partial_{\delta} - \chi_{\alpha \beta}^{(2,0)} \partial_{t}^2 \biggr ] n_\beta(\rr, t).
\end{align}
The equation of motion takes the form $\dot{\nn} = \nn \cross \vb{L}$, where the vector $\vb{L} = \tilde{\vb{L}}+\vb{L}^{(2)}$ consists of various terms. $\tilde{\vb{L}}$ contains the intrinsic exchange interaction $J_\text{ex}$ and easy-axis anisotropy $K$. It can also contain Gilbert damping and external fields, but these are left out here for simplicity. The FM's intrinsic part $\tilde{\vb{L}}$ is shown in Appendix \ref{sec:llg_appendix}. The term of interest, $\vb{L}^{(2)}$, contains the induced interactions mediated by the TI fermions. This term can be found using the Euler-Lagrange equations and has the general form
\begin{align}\nonumber
    &L^{(2)}_\alpha =\frac{1}{2\pi} \biggl [ \left( \chi_{\alpha \beta}^{(0,0)} + \chi_{\beta \alpha }^{(0,0)} \right)+ i \left(\chi_{\alpha \beta \gamma}^{(0,1)} - \chi_{\beta \alpha \gamma}^{(0,1)} \right)\partial_{\gamma}
     \\ \label{eq:llg_L_general}
     &- i \left( \chi_{\alpha \beta }^{(1,0)} - \chi_{\beta \alpha }^{(1,0)} \right) \partial_{t}+ \left(\chi_{\alpha \beta \gamma}^{(1,1)} + \chi_{\beta \alpha \gamma}^{(1,1)}\right)\partial_{t} \partial_{\gamma} \\\nonumber
     & - \left ( \chi_{\alpha \beta \gamma \delta}^{(0,2)} + \chi_{\beta \alpha \gamma \delta}^{(0,2)} \right) \partial_{\gamma} \partial_{\delta}
     - \left( \chi_{\alpha \beta}^{(2,0)} + \chi_{\beta \alpha}^{(2,0)} \right) \partial_{t}^2 \biggr ] n_\beta(\rr, t).
\end{align}
From Eq. \eqref{eq:llg_L_general}, it follows that terms with an \emph{even} number of derivatives can originate only from the spin-symmetric parts of the exchange tensor $\chi_{\alpha \beta}$. Similarly, the DM-like interactions, which are $\chi_{\alpha \beta}$'s antisymmetric part, only appear in terms with an odd number of derivatives, i.e. the $\partial_\gamma$ and $\partial_t$ terms.

\begin{figure*}
    \centering
    \begin{subfigure}[b]{0.45\textwidth}
        \centering
        \includegraphics[width=\linewidth]{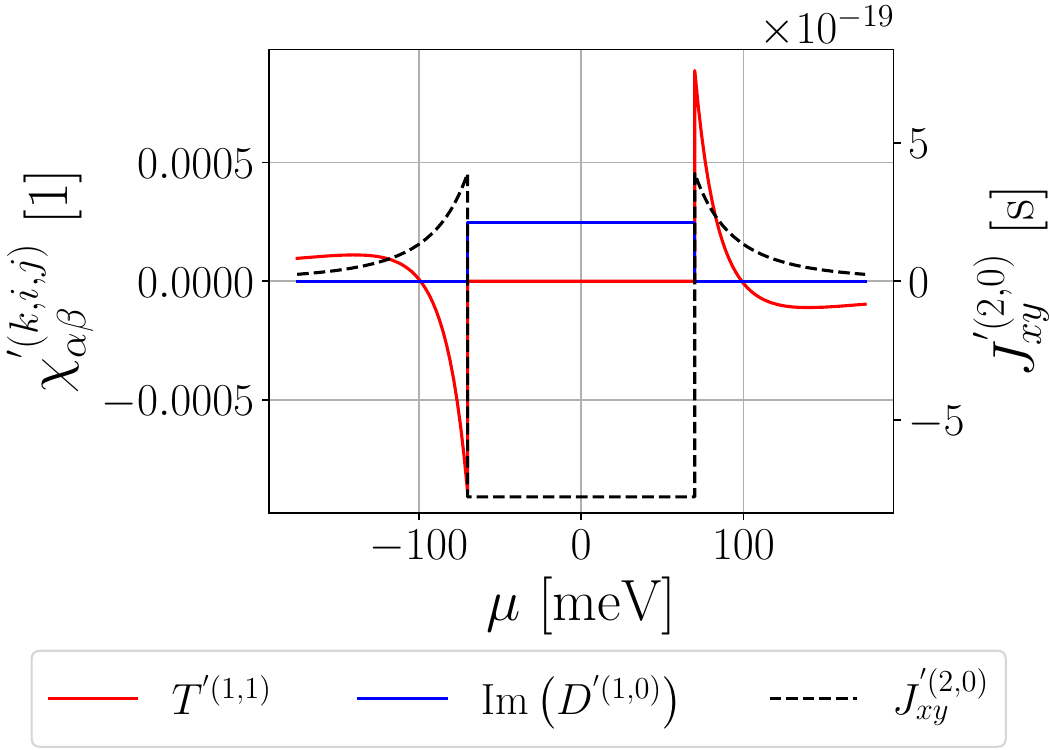}
    \end{subfigure}
    \begin{subfigure}[b]{0.49\textwidth}
        \centering
        \includegraphics[width=\linewidth]{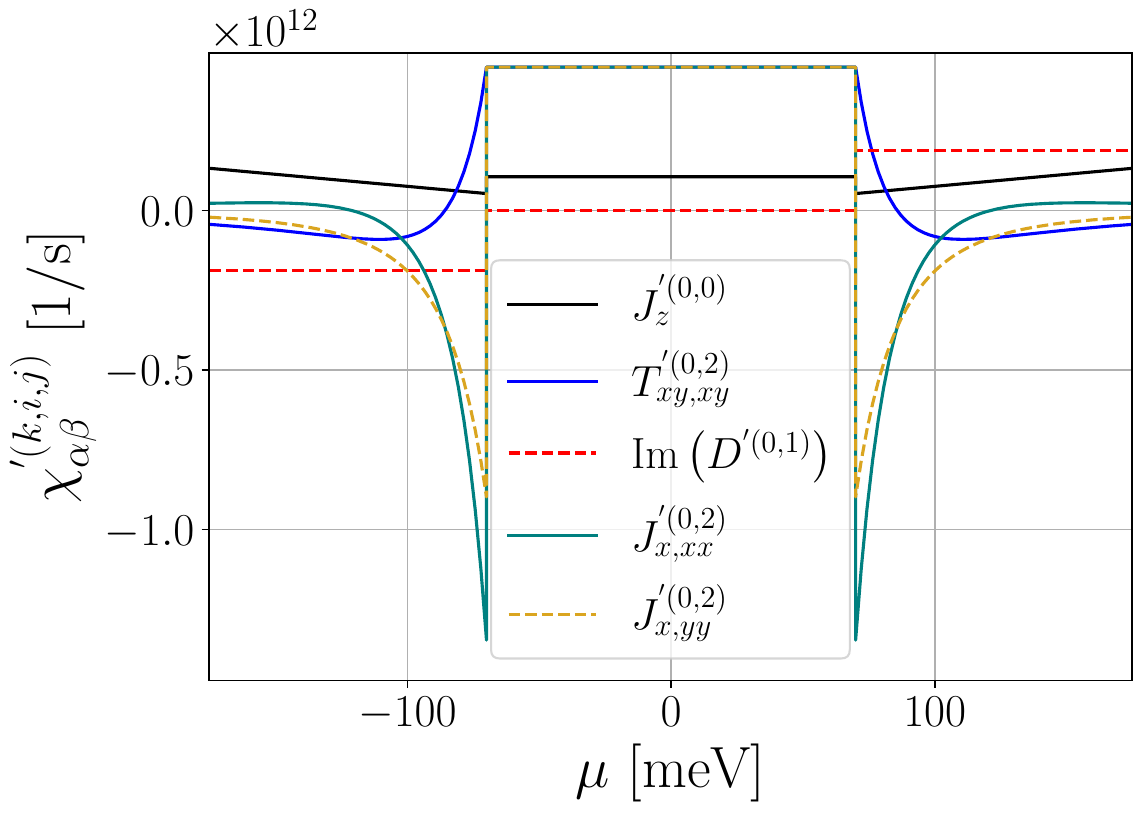}
    \end{subfigure}
    \caption{\waschanged{\small Spin-spin interaction coefficients appearing in Eq.~\eqref{eq:H_eff_prime_form} in the main text as a function of chemical potential $\mu$. All coefficients are constant for $\mu$ in the gap, $m_0=2\bar{J}$, given by the interfacial exchange interaction $\bar{J}=\SI{35}{\milli \electronvolt}.$ The DMI term $D^{(0,1)}$ and the skyrmion-dynamics term $T^{(1,1)}$ display a switching effect, where they suddenly become zero whenever $\abs{\mu} < m_0.$}}
    \label{fig:interactions}
\end{figure*}

The structure of the equation of motion is
\begin{align}
    \dot{\nn} &= \nn \cross \vb{H}_\text{eff}^\prime,
\end{align}
with
\begin{widetext}
    \begin{align}\label{eq:H_eff_prime_form}
    \vb{H}_\text{eff}^\prime &= 
    \begin{pmatrix}
        J_{xy}^{\prime(0,0)} n_x\\
        J_{xy}^{\prime(0,0)} n_y \\
        J_z^{\prime(0,0)} n_z
    \end{pmatrix} - T_{xy,xy}^{\prime(0,2)}\begin{pmatrix}
        \partial_{xy} n_y\\
        \partial_{xy} n_x\\
        0
    \end{pmatrix}
     - \begin{pmatrix}
        J_{xy}^{\prime(2,0)} \ddot{n}_x \\
        J_{xy}^{\prime(2,0)} \ddot{n}_y\\
        J_z^{\prime(2,0)} \ddot{n}_z
    \end{pmatrix} - \begin{pmatrix}
        \left ( J_{x, xx}^{\prime(0,2)}\partial_x^2 +J_{x, yy}^{\prime(0,2)}\partial_y^2 \right) n_x\\
        \left ( J_{x, yy}^{\prime(0,2)}\partial_x^2 +J_{x, xx}^{\prime(0,2)}\partial_y^2 \right) n_y\\
        J_{z, xx}^{\prime(0,2)}\left ( \partial_x^2 +\partial_y^2 \right) n_z
    \end{pmatrix}  \\ \nonumber
    &\phantom{QQQ} +  T^{\prime(1,1)}\begin{pmatrix}
        \waschanged{(\grad \cross \dot{\nn})_x} \\
        \waschanged{(\grad \cross \dot{\nn})_y} \\
        \waschanged{-(\grad \cross \dot{\nn})_z}
    \end{pmatrix}
    + i D^{\prime(0,1)} \begin{pmatrix}
        - \partial_x n_\waschanged{z}\\
        -\partial_y n_z\\
        \partial_y n_y + \partial_x n_x
    \end{pmatrix} + i D^{\prime(1,0)} \hat{\vb{z}} \cross \dot
    {\nn} + \tilde{\vb{L}},
\end{align}
\end{widetext}
where some terms are exclusively present when $\mu$ is either inside or outside the gap. The full overview is shown at the end of Appendix \ref{sec:dynamic_appendix}. The primed quantities are a rescaled version of the unprimed ones, $\chi_{\alpha \beta}^{\prime} = \chi_{\alpha\beta} / \pi.$ \waschanged{In Fig.~\ref{fig:interactions}, a selection of the above interaction coefficients are shown as a function of chemical potential $\mu$. Nearly all the coefficients change character or go from zero to finite as $\mu$ passes through the values $\mu=\pm m_0.$}

It can be shown that the inclusion of the $D^{\prime(1,0)}$ term, which is finite only for $\abs{\mu} < m_0$, leads to a slight renormalization of all parameters, including potential Gilbert damping and nutation terms, by a factor $1/(1 + iD^{\prime(1,0)} n_z)$, relevant for ferromagnetic resonance experiments. 

Equation \eqref{eq:H_eff_prime_form} contains the terms $T^{(0,2)}$ and $D^{(0,1)}$ relevant to skyrmions, as reported elsewhere \cite{Rex2016,Nogueira2018}. The extension to a dynamical theory reveals a term $T^{(1,1)}$ that -- to our knowledge -- is new in the setting of spin interactions mediated by TI fermions. It has both time and space derivatives, which means it should be present in a dynamical nonuniform spin texture. It is in fact composed of the components of $\grad \cross \dot{\nn}$ and may therefore be relevant for e.g. skyrmion dynamics\waschanged{.} Although a full solution of the LLG equation in such a case is beyond the scope of this work\waschanged{, qualitative statements about the effective field arising from the $T^{(1,1)}$ term can be made. We first consider a Néel skyrmion, which has $\grad\cross \nn$ entirely in-plane. If the skyrmion is in its breathing mode due to e.g. thermal fluctuations or external-field fluctuations, then $\partial_t \grad\cross \nn$ is finite and in-plane since the spatial gradients change with changing skyrmion radius. For a Bloch skyrmion, with finite in-plane rotation, there will also be a $z$ component to $\partial_t \grad \cross \nn$. Depending on specifics of the skyrmion profile, its breathing mode and potentially where in the skyrmion, the $x$ and $y$ components will either counteract or accelerate the breathing motion. Due to the minus sign in Eq.\ \eqref{eq:H_eff_prime_form}, the $z$ component will have the opposite effect, which may lead to interesting behavior for the stability of Bloch skyrmions, where all components are expected to be operative and have effects that are in tension with one another.} Note that the new term is present only for $\abs{\mu} > m_0$, which explains why studying magnetization dynamics at $\mu=0$ did not reveal it \cite{Rex2016}.

\waschanged{To obtain some more specific predictions,} we instead show how this theory applies to linear spin waves around the near-uniform state in the next subsection\waschanged{, where we find that the magnon gaps are tunable by tuning $\mu$, and where a softening of the high-frequency inertial spin-wave mode is predicted.}

\subsection{Linear response}
\begin{figure*}
    \centering
    \begin{subfigure}[b]{0.32\textwidth}
        \centering
        \includegraphics[width=\linewidth]{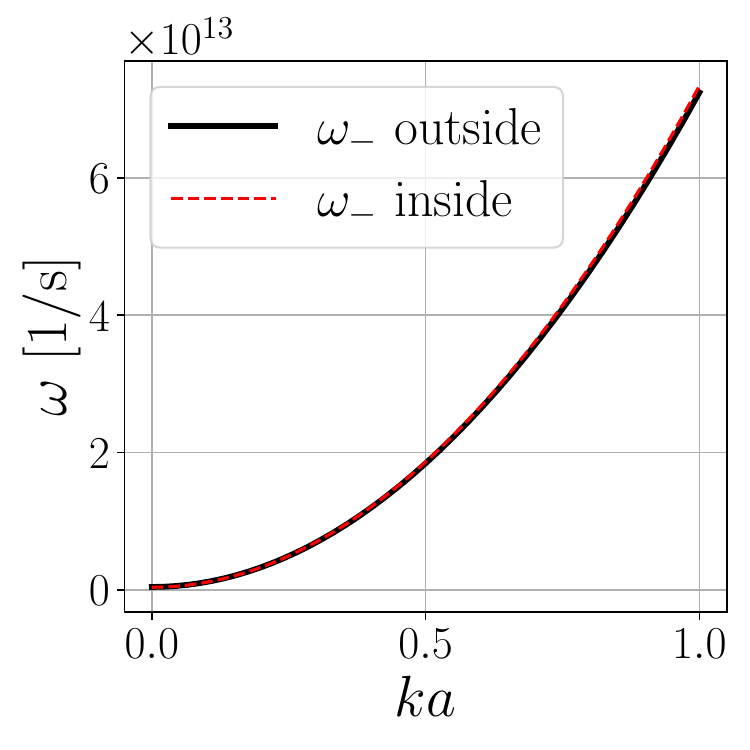}
    \end{subfigure}
    \begin{subfigure}[b]{0.32\textwidth}
        \centering
        \includegraphics[width=\linewidth]{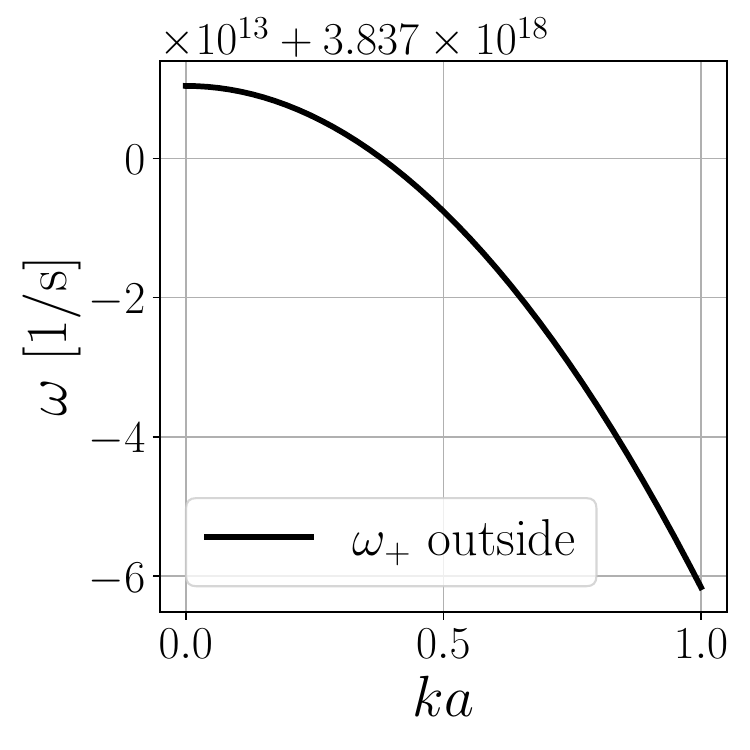}
    \end{subfigure}
    \begin{subfigure}[b]{0.32\textwidth}
        \centering
        \includegraphics[width=\linewidth]{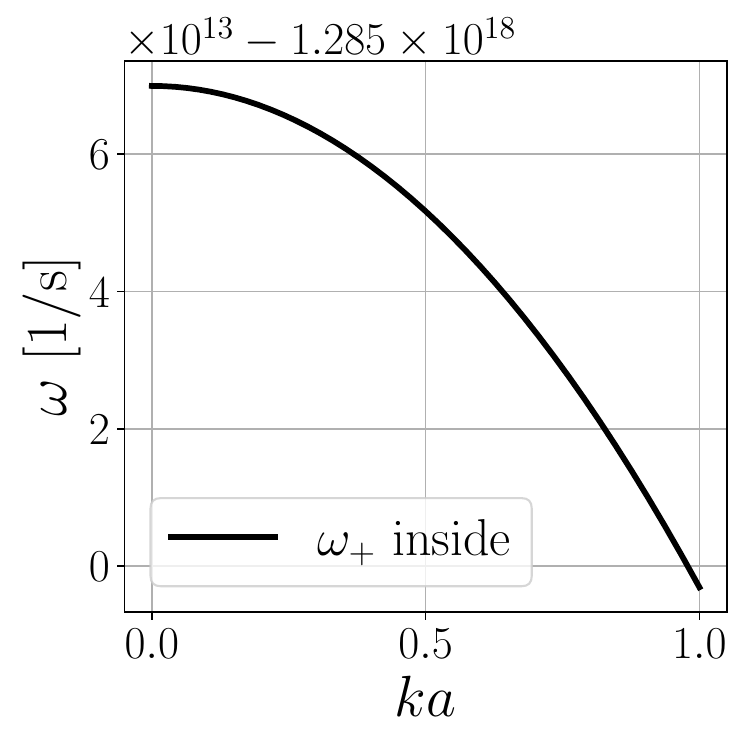}
    \end{subfigure}
    \caption{\small Spin-wave dispersion relations for the precession modes $\omega_-$ and the nutation modes $\omega_+$ using chemical potential $\mu=\SI{80}{\milli \electronvolt}$ for the outside-the-gap curves, interfacial exchange $\bar{J}=\SI{35}{\milli \electronvolt}$, and Fermi velocity $\hbar \vF/a=\SI{500}{\milli \electronvolt}$. The FM's intrinsic easy-axis anisotropy was set to $K=\SI{1}{\milli \electronvolt}$, and the intrinsic exchange interaction to $J_\text{ex}=\SI{150}{\milli \electronvolt}$. Further analysis and interpretations of negative frequencies are explained in the main text.}
    \label{fig:omegas_compareInVsOut}
\end{figure*}

One way to see the effect of some of the many new terms generated by the TI, which involve both time derivatives and spatial gradients, is to study the dispersion $\omega(\kk)$ of linear spin waves. We explain the calculations in Appendix \ref{sec:llg_appendix}, where we use the approximation
\begin{align}
    \nn(\rr, t) = \tilde{m}_0 \hat{\vb{z}} + \epsilon \vb{s}\e{i(\kk \cdot \rr - \omega t)}
\end{align}
and expand the LLG equation to first order in the smallness parameter $\epsilon$. It yields a linearized LLG equation for the spin fluctuation amplitudes $\vb{s}$. Searching for a solution with $s_x=\e{i\phi}s_y > 0$ for real phase difference $\phi$, we then obtain an expression for $\omega(\kk)$ that is isotropic. The solutions are only stable for $\phi = \pm \pi/2$, i.e. for circular motion about a fixed $n_z=\tilde{m}_0.$ Choosing $\phi=-\pi/2$, the dispersion relation reads
\begin{align}
    \label{eq:disp_rel}
    \omega_\pm(\kk) & = \frac{1}{2J^{\prime(2,0)}_{xy}} \left(1+iD^{\prime(1,0)} \pm \sqrt{R} \right) \mp \frac{J_\text{eff}}{2\sqrt{R}} k^2.
\end{align}
Here, we defined
\begin{subequations}\begin{align}
    R &= \left(1+iD^{\prime(1,0)}\right)^2 - 4J_{xy}^{\prime (2,0)} K_\text{eff} \\
    J_\text{eff} &= J^{\prime(0,2)}_{x,xx} + J^{\prime(0,2)}_{x,yy} + 2 \frac{J_\text{ex}}{\pi}\\
    K_\text{eff} &= \frac{K}{\pi} - J_z^{\prime(0,0)} + J_{xy}^{\prime(0,0)}.
\end{align}
\end{subequations}
Terms such as $T^{(1,1)}, T^{(0,2)}$, and $D^{(0,1)}$, relevant for more nonuniform spin textures like skyrmions, cancel or drop out in this consideration of small perturbations around the uniform state. The dispersion relation is shown in Fig.\ \ref{fig:omegas_compareInVsOut} for a $\mu$ slightly outside the gap $m_0 = 2\bar{J}=\SI{70}{\milli \electronvolt}$ and for $\mu$ inside the gap, where all the induced interactions are constants as functions of $\mu$. 

To first order in the small dimensionless quantity $R-1$, we have
\begin{align}
    \omega_+(0) &\approx  \frac{\left(1+\frac{iD^{\prime(1,0)}}{2}\right)^2 }{J_{xy}^{\prime(2,0)}} - K_\text{eff}\\
     \omega_-(0) &\approx \frac{\left(D^{\prime(1,0)}\right)^2}{4 J_{xy}^{\prime (2,0)}} + K_\text{eff}.
\end{align}
The precession branch has finite $\omega_-(0)$ because of the easy-axis anisotropy in the FM, which is renormalized by the spin \emph{anisotropic} $J_\alpha^{(0,0)}$ term, leading to the effective easy-axis anisotropy $K_\text{eff}$. This anisotropy is the dominant term in $\omega_-(0)$. The dominant term in $\omega_+(0)$, however, is the $1/J_{xy}^{\prime (2,0)}$ term. The dispersion is renormalized in somewhat different ways when $\mu$ is inside and outside the gap, as can be seen in Fig.\ \ref{fig:omegas_compareInVsOut}. The largest difference is the switching sign of the nutation term $J_{xy}^{(2,0)},$ which essentially causes $\omega_+(k)$ to curve downwards for $\mu$ outside the gap, no longer favoring the uniform mode $k=0.$ 

The above statement may seem wrong since the $k^2$ coefficient of $\omega_+(\kk)$ is independent of the nutation term. Indeed, both nutation branches curve downward, but their $\omega_+(0)$ values have different sign because of the nutation term's sign swap as $\mu$ is tuned to lie inside or outside the gap. As noted by \waschanged{Kikuchi} et al. \cite{Kikuchi2015}, the all-negative nutation branch $\omega_+$, which is for chemical potential $\abs{\mu} < m_0$ inside the gap, should be viewed in the context of a superposition of the nutation mode on top of the precession mode. The precession mode for $\abs{\mu} < m_0$, then, takes the magnetization counterclockwise around the effective field at a frequency $\omega_-$, and on top of that, there is a small nutation mode with frequency $-\omega_+>0$ in the clockwise direction. For $\mu$ outside the gap, $\omega_+(0)>0$ because the nutation term is positive, and the graph indeed curves downward, but stays positive, displaying softening of the nonuniform nutation modes $k>0.$
\ \\
\ \\
\section{Conclusion}
In this paper we have presented an analysis of magnetic heterostructures composed of ferromagnetic layers interfaced with three-dimensional topological insulators (TIs). We have studied the TI-induced two-spin interactions, which take the form of generalized RKKY exchange interactions. The interplay between the spin-orbit coupling intrinsic to the TI and the magnetization in the FM layer leads to highly nonlocal and retarded, chiral, and Dzyaloshinskii-Moriya-like contributions to the effective spin Hamiltonian. In the static case, where interactions are instantaneous, the relevant length scales of the spatial oscillations are given by the Fermi momentum $\kF$, and the chemical potential $\mu$ and the interfacial exchange strength $\bar{J}$ set the amplitude of the oscillations. These parameters, then, allow for tuning of how much the TI electrons affect the magnetization texture. Furthermore, we have studied the spin dynamics through a derivation of the LLG equation. In the general case, we have found that the LLG equation's parameters are renormalized by the interface with the TI, but a term related to the time derivative of the curl of the magnetization, relevant for skyrmion dynamics, also appears if the chemical potential is above a certain threshold. The magnon dispersion is modified by the TI-mediated interactions, including tunable magnon gaps, sensitive to a tunable chemical potential and interfacial exchange coupling strength. The nutation term, related to the second time derivative of the magnetization, switches signs as the chemical potential is tuned outside the gap in the electron dispersion, softening the nonuniform high-frequency magnon modes.

\acknowledgements
This work was supported by the Research Council of Norway (RCN) through its Centres of Excellence funding scheme, Project No. 262633, ``QuSpin'', RCN Project No. 323766, as well as COST Action CA21144  ``Superconducting Nanodevices and Quantum Materials for Coherent Manipulation".

\bibliography{main.bib}

\begin{widetext}
\appendix
\section{Integrating out Dirac fermions}
\label{sec:path_integrals}
We begin by going to reciprocal space in the full action from Eq.\ \eqref{eq:partition_function} in the main text using the Fourier transform conventions
\begin{subequations}
\begin{align}
    &\psi(\rr,\tau) = \frac{1}{\beta} \sum_{n} \int \frac{\dd[2]{k}}{(2\pi)^2} \psi_{n}(\kk) \e{i(\kk\cdot \rr - \omega_n \tau)} \label{eq:FT_psi_general_forward}\\
    & \psi_{n}(\kk) = \int_0^\beta \dd{ \tau} \int \dd[2]{r} \psi(\rr, \tau) \e{-i(\kk\cdot\rr  - \omega_n \tau)},\label{eq:FT_psi_general_backward}
\end{align}
\end{subequations}

where $\rr, \kk$ are dimensionless, both scaled by the lattice constant, $a$ throughout the paper. Then,
\begin{align}
    \psi^\dagger \left ( G^{-1} + B \right) \psi &= \frac{1}{\beta^2} \int \frac{\dd[2]{k_1}}{(2\pi)^2}\int\frac{\dd[2]{k_2}}{(2\pi)^2} \sum_{n_1, n_2} \psi^\dagger_{n_1}(\kk_1) \int\dd[2]{r} \dd{\tau} \e{-i(\kk_1\cdot \rr - \omega_{n_1}\tau)} \\\nonumber
    &\times \left[ \partial_\tau + \begin{pmatrix}
        -i\hbar \vF \partial_y\\
        i\hbar \vF \partial_x\\
        -m_0
    \end{pmatrix} \cdot \bsigma - \mu + \frac{1}{\beta} \sum_{n_3} \int \frac{\dd[2]{k_3}}{(2\pi)^2} B(\nn(\kk_3, i\omega_{n_3})) \e{i(\kk_3 \cdot \rr - \omega_{n_3} \tau)}\right] \e{i(\kk_2 \cdot \rr - \omega_{n_2}\tau)} \psi_{n_2}(\kk_2)
\end{align}
for the choice $\vb{d}_1$ of SOC, Eq. \eqref{eq:d1}. For the other choice, Eq. \eqref{eq:d2}, one proceeds in the same way, so we omit this calculation for brevity. Here,
\begin{align}
        \int\dd[2]{r} \dd{\tau} \e{i(\rr\cdot \kk - \tau \omega_n)} = \beta (2\pi)^2 \delta(\kk)\delta_{n,0},
\end{align}
so
\begin{align}
    \psi^\dagger \left ( G^{-1} + B \right) \psi &= \int \frac{\dd[2]{k_1}}{(2\pi)^2}\int\frac{\dd[2]{k_2}}{(2\pi)^2} \sum_{n_1, n_2}  \psi^\dagger_{n_1}(\kk_1) \left ( G_{\kk_1, \kk_2; n_1, n_2}^{-1} + B_{\kk_1, \kk_2; n_1, n_2} \right) \psi_{n_2}(\kk_2),
\end{align}
where
\begin{subequations}
\begin{align}
    G_{\kk_1, \kk_2; n_1, n_2}^{-1} &= \frac{(2\pi)^2}{\beta} \left [-i\omega_{n_2} + \vb{d}(\kk_2)\cdot\bsigma - \mu \right]\delta_{n_1, n_2} \delta(\kk_1 - \kk_2)\\
    B_{\kk_1, \kk_2; n_1, n_2} &= \frac{-2\bar{J}}{\beta^2} \bsigma \cdot \nn(\kk_1 - \kk_2, i\omega_{n_1} - i\omega_{n_2}).
\end{align}
\end{subequations}
Now we integrate out the fermions $\psi$,
\begin{align}
    \int \mathcal{D}\psi \mathcal{D}\psi^\dagger \e{-\psi^\dagger \left (G^{-1} + B\right)\psi} &= \e{\Tr \ln \left ( G^{-1} + B \right)} \equiv \e{-\Delta S_\text{eff}[\nn]},
\end{align}
such that the effective action reads
\begin{align}
    S_\text{eff}[\nn] = S_\text{FM}[\nn] + \Delta S_\text{eff}[\nn]
\end{align}
with $\Delta S_\text{eff}[\nn] = - \Tr \ln \left ( G^{-1} + B \right)$, which can be written
\begin{subequations}
    \begin{align}
     \Delta S_\text{eff}[\nn]&= -\Tr \ln \left ( G^{-1} (\mathbb{I} + GB) \right) \\
     &\approx - \Tr \ln G^{-1} \underbrace{- \Tr (GB)}_{\equiv \Delta S_\text{eff}^{(1)}[\nn]} + \underbrace{\frac{1}{2} \Tr (GBGB)}_{\equiv \Delta S_\text{eff}^{(2)}[\nn]}.
\end{align}
\end{subequations}
Here, we expanded to second order in the spin-fermion interaction vertex. The electron propagator is given by the inverse of $G^{-1}$, i.e.,
\begin{align}
    G =\frac{\beta}{(2\pi)^2} \delta_{n_1, n_2} \delta(\kk_1 - \kk_2) \frac{-\vb{d}(\kk_2) \cdot \bsigma - d_0(i\omega_{n_2})\mathbb{I} }{(i\omega_{n_2} + \mu)^2 - d_x^2 - d_y^2 - d_z^2},
\end{align}
where $d_0(i\omega_{n}) \equiv i\omega_{n} + \mu$. 

The trace over all quantum numbers, $\Tr$, can be simplified to a trace $\tr$ in spin space,
\begin{subequations}
    \begin{align}
        \Delta S_\text{eff}^{(2)}[\nn] &= \frac{(-2\bar{J})^2}{2\beta^2} \sum_{n_1,n_2}  \int \frac{\dd[2]{k_1}}{(2\pi)^2}\frac{\dd[2]{k_2}}{(2\pi)^2} \frac{1}{\left [(i\omega_{n_1} + \mu)^2 - d^2(\kk_1) \right]\left [(i\omega_{n_2} + \mu)^2 - d^2(\kk_2) \right]} \\\nonumber
    &\times \tr \left [ \bigl (\sigma_\gamma d_\gamma(\kk_1) + d_0(i\omega_{n_1}) \mathbb{I} \bigr) \sigma_\alpha n_\alpha(k_1 - k_2) \bigl(\sigma_\delta d_\delta(\kk_2)+ d_0(i\omega_{n_2})\mathbb{I}\bigr) \sigma_\beta  n_\beta(k_2 - k_1) \right].
    \end{align}
\end{subequations}
The presence of SOC in the TI adds more off-diagonal factors $\sigma_\gamma d_\gamma(\kk)$ in the above spin trace than would be generated by normal-metal electrons with spin-split parabolic bands \cite{Johnsen2025}. This, in turn, will allow spin-spin interactions that are off-diagonal, which disappear due to inversion symmetry in the normal-metal case. The spin trace above can be performed using the Pauli matrix properties
\begin{subequations}
\begin{align}
    \tr \sigma_\alpha \sigma_\beta &= 2 \delta_{\alpha \beta}\\
    \tr \sigma_\alpha \sigma_\beta \sigma_\gamma &= 2i \epsilon_{\alpha \beta \gamma}\\
    \tr \sigma_\alpha \sigma_\beta \sigma_\gamma \sigma_\delta &= 2 ( \delta_{\alpha \beta}\delta_{\gamma \delta} - \delta_{\alpha\gamma} \delta_{\beta \delta} + \delta_{\alpha \delta}\delta_{\beta \gamma} )
\end{align}
\end{subequations}
to yield
\begin{align}\nonumber
    \Delta S_\text{eff}^{(2)}[\nn] &= \frac{4\bar{J}^2}{\beta^2} \sum_{n_1, n_2} \int\frac{\dd[2]{k_1}}{(2\pi)^2}\frac{\dd[2]{k_2}}{(2\pi)^2}\frac{n_\alpha(k_1 - k_2) n_\beta(k_2 - k_1)}{\left [(i\omega_{n_1} + \mu)^2 - d^2(\kk_1) \right]\left [(i\omega_{n_2} + \mu)^2 - d^2(\kk_2) \right]}\\
    &\phantom{QQQQQ} \times \biggl [ d_\gamma(\kk_1)d_\delta(\kk_2) (\delta_{\gamma \alpha} \delta_{\delta \beta} - \delta_{\gamma \delta} \delta_{\alpha \beta} + \delta_{\gamma \beta} \delta_{\alpha \delta}) + d_\gamma(\kk_1) d_0(i\omega_{n_2}) i \epsilon_{\gamma \alpha \beta} \\\nonumber
    &\phantom{QQQQQQQ} + d_0(i\omega_{n_1}) d_\gamma(\kk_2) i\epsilon_{\alpha \gamma \beta} + d_0(i\omega_{n_1}) d_0(i\omega_{n_2}) \delta_{\alpha \beta} \biggr].
\end{align}
Next, straightforward manipulations permit us to rewrite the four-Pauli-matrix term to
\begin{align}
    d_\gamma(\kk_1) d_\delta(\kk_2)(\delta_{\gamma \alpha} \delta_{\delta \beta} - \delta_{\gamma \delta} \delta_{\alpha \beta} + \delta_{\gamma \beta} \delta_{\alpha \delta}) &= \bigl ( d_\alpha(\kk_1) d_\beta(\kk_2) + d_\beta(\kk_1) d_\alpha(\kk_2) \bigr)  (1 - \delta_{\alpha \beta}) \\\nonumber
    & + \biggl ( d_\alpha(\kk_1) d_\alpha(\kk_2) - \sum_{\gamma \neq \alpha} d_\gamma(\kk_1) d_\gamma(\kk_2)  \biggr)\delta_{\alpha \beta},
\end{align}
showing that it is comprised of a spin-diagonal part and an off-diagonal symmetric part. Collecting the terms with similar spin structure, we obtain
\begin{align}
    \Delta S_\text{eff}^{(2)}[\nn] &= \frac{4\bar{J}^2}{\beta^2} \sum_{n_1, n_2} \int\frac{\dd[2]{k_1}}{(2\pi)^2}\frac{\dd[2]{k_2}}{(2\pi)^2}\frac{n_\alpha(\kk_1 - \kk_2, i\omega_{n_1} - i\omega_{n_2}) n_\beta(\kk_2 - \kk_1, i\omega_{n_2} - i\omega_{n_1})}{\left [(i\omega_{n_1} + \mu)^2 - d^2(\kk_1) \right]\left [(i\omega_{n_2} + \mu)^2 - d^2(\kk_2) \right]}\\ \nonumber
    &\times \Biggl [ \biggl(d_0(i\omega_{n_1}) d_0(i \omega_{n_2})+d_\alpha(\kk_1) d_\alpha(\kk_2) - \sum_{\gamma \neq \alpha} d_\gamma(\kk_1) d_\gamma(\kk_2)\biggr) \delta_{\alpha \beta} \\\nonumber
    & + i \Bigl( d_\gamma(\kk_1) d_0(i\omega_{n_2}) - d_\gamma(\kk_2) d_0(i\omega_{n_1}) \Bigr) \epsilon_{\alpha \beta \gamma}   + \Bigl ( d_\alpha(\kk_1) d_\beta(\kk_2) + d_\beta(\kk_1) d_\alpha(\kk_2) \Bigr)  (1 - \delta_{\alpha \beta}) \Biggr],
\end{align}
i.e., Eq.\ \eqref{eq:DTJ_general_start} in the main text. The first term represents a generated anisotropic exchange term (a generalized RKKY interaction). The second is a generated DMI interaction term, and the third term is an off-diagonal exchange term that couples different components of the spins. The two latter both originate with the SOC of the TI, whereas the first term would have a non-zero counterpart if the TI were replaced by a NM with no SOC.   

\section{Momentum integrals and Matsubara sums for the static theory}
\label{sec:static_appendix}
Here we show how to compute 
\begin{align}\nonumber
    &\chi_{\alpha \beta}(\Delta \rr, i\omega_\nu) = \frac{1}{\beta}\sum_n\int \frac{\dd[2]{k_1}}{(2\pi)^2} \int\frac{\dd[2]{k_2}}{(2\pi)^2} \e{-i(\kk_1-\kk_2)\cdot \Delta \rr} \biggl[\delta_{\alpha \beta} J_\alpha(\kk_1, \kk_2, i\omega_n + i\omega_\nu, i\omega_n)  \\
    &  + (1-\delta_{\alpha\beta}) T_{\alpha \beta}^{\text{sym}}(\kk_1, \kk_2, i\omega_n + i\omega_\nu, i\omega_n)+\epsilon_{\alpha \beta \gamma} D_\gamma(\kk_1, \kk_2, i\omega_n + i\omega_\nu, i\omega_n)\biggr]
\end{align}
in the static limit, $i\omega_\nu = 0$. We treat each term $J_\alpha, D_\gamma, T_{\alpha \beta}^\text{sym}$ separately because they are sums of separable $\kk$ integrals and one Matsubara sum, as seen from Eq.\ \eqref{eq:DTJ_general_start}. Starting with the only term where we can obtain an analytic expression at zero temperature, $D_\gamma$, we note that since
\begin{align}\label{eq:D_gamma_static_start}
    &D_\gamma(\Delta \rr, i\omega_\nu = 0) = \frac{4i \bar{J}^2 }{\beta}\int \frac{\dd[2]{k_1}}{(2\pi)^2} \int\frac{\dd[2]{k_2}}{(2\pi)^2}  \sum_n [d_\gamma(\kk_1)- d_\gamma(\kk_2)] \frac{X_n\e{-i(\kk_1-\kk_2)\cdot \Delta \rr}}{\Bigl[X_n^2 - d^2(\kk_1)\Bigr]\Bigl[X_n^2 - d^2(\kk_2)\Bigr]},
\end{align}
we have $D_z(\Delta \rr, i\omega_\nu=0) = 0$ on account of $d_z(\kk) = -m_0$ being $\kk$ independent. This result does not depend on the choice of $\vb{d}$ vector, Eqs. \eqref{eq:d1} or \eqref{eq:d2}. It relies only on the fact that $d_x,d_y$ are odd in momentum and that $d_z$ is momentum independent since the TI surface states have no dispersion perpendicular to the surface. Here, $X_n \equiv i\omega_n + \mu$. Assuming $\gamma \neq z$, the above can be simplified to 
\begin{align}
    &D_\gamma(\Delta \rr, i\omega_\nu=0) = \frac{-8 \bar{J}^2 }{\beta}\Im \sum_n \int\frac{\dd[2]{k_1}}{(2\pi)^2} \frac{\dd[2]{k_2}}{(2\pi)^2} \frac{d_\gamma(\kk_1) X_n \e{-i(\kk_1-\kk_2)\cdot \Delta \rr}}{\Bigl[X_n^2 - d^2(\kk_1)\Bigr]\Bigl[X_n^2 - d^2(\kk_2)\Bigr]}
\end{align}
as can be seen by swapping $\kk_1 \leftrightarrow \kk_2$ and $i\omega_n \to -i \omega_n$ in the $d_\gamma(\kk_2)$ term. 
The DMI term is separable in momenta and given by 
\begin{align}
    D_\gamma(\Delta \rr, i\omega_\nu=0) = \frac{-8\bar{J}^2}{\beta} \Im \sum_n (i\omega_n+\mu)V_\gamma(-\Delta \rr, i\omega_n)U(\Delta \rr, i\omega_n),
\end{align}
where
\begin{subequations}
\begin{align}
    U(\Delta \rr, i\omega_n) &= \int\frac{\dd[2]{k}}{(2\pi)^2}\frac{\e{i\kk\cdot\Delta \rr}}{(i\omega_n + \mu)^2 - d^2(\kk)}\\ \label{eq:def_V_gamma}
    V_\gamma(\Delta \rr, i\omega_n) &= \int\frac{\dd[2]{k}}{(2\pi)^2}\frac{d_\gamma(\kk)\e{i\kk\cdot\Delta \rr}}{(i\omega_n + \mu)^2 - d^2(\kk)}.
\end{align}
\end{subequations}
Note that $V_{\gamma \neq z}$ can be obtained from $U$ by differentiation and that $V_z = -m_0 U$.

The basic momentum integral needed to find $\chi_{\alpha\beta}(\Delta \rr, i\omega_\nu=0),$ then, is
\begin{align}
    U(\Delta \rr, i\omega_n) &= \frac{1}{2\pi} \int_0^\infty \dd{k} k\int_0^{2\pi}\frac{\dd{\phi}}{2\pi} \frac{\e{i\abs{\kk} \abs{\Delta \rr} \cos \phi}}{(i\omega_n + \mu)^2 - d^2(k)}\\
    &= \frac{1}{2\pi} \int_0^\infty \dd{k} k \frac{J_0(\abs{\Delta \rr} k)}{(i\omega_n + \mu)^2 - d^2(k)}
\end{align}
since the dispersion $d(\kk)=d(k) = (\hbar \vF k)^2 + m_0^2$ is isotropic. Therefore,
\begin{align}
    U(\Delta \rr, i\omega_n) &= \frac{-1}{2\pi(\hbar \vF)^2}\int_0^\infty \dd{k} k \frac{J_0(\abs{\Delta \rr} k)}{k^2 + \frac{m_0^2 - (i\omega_n + \mu)^2}{(\hbar \vF)^2}},
\end{align}
so $U(\Delta \rr, i\omega_n)$ is of the form \cite{GnR}
\begin{align}
    \int_0^\infty \dd{k} k \frac{J_0(ak)}{k^2 + b^2} = K_0(ab),
\end{align}
which holds for $a > 0, \Re b > 0$. Thus, we choose the principal branch of the square root in
\begin{align}
    b = \sqrt{\frac{m_0^2 - (i\omega_n + \mu)^2}{(\hbar \vF)^2}},
\end{align}
i.e., $\sqrt{\abs{z}\e{i\theta}} = \sqrt{\abs{z}} \e{i\theta/2}$ with the phase $\theta \in (-\pi, \pi].$ Using the result
\begin{align}
    U(\Delta \rr, i\omega_n) &= \frac{-1}{2\pi (\hbar \vF)^2} K_0\left( \frac{\abs{\Delta \rr}}{\hbar \vF} \sqrt{m_0^2 - (i\omega_n +\mu)^2} \right),
\end{align}
the expression for the other $\kk$ integral follows by differentiation:
\begin{align}
    V_{\gamma \neq z}(\Delta \rr, i\omega_n) &= \frac{i}{2\pi (\hbar \vF)^2}\frac{\sqrt{m_0^2-(i\omega_n + \mu)^2}}{ \abs{\Delta \rr}}\begin{pmatrix}
         -\Delta y\\\Delta x
    \end{pmatrix}_\gamma K_1\left (\frac{\abs{\Delta \rr}}{\hbar \vF}\sqrt{\left [m_0^2 - (i\omega_n + \mu)^2\right] } \right).
\end{align}
This can now be inserted into $D_\gamma$ to perform the Matsubara summation, but let us first show how $J_\alpha$ and $T_{\alpha \beta}^\text{sym}$ can be expressed in terms of the same $U, V_\gamma$.

From the basic expression
\begin{align}\label{eq:T_static_starting}
    T_{\alpha \beta}^{\text{sym}}(\Delta \rr, i\omega_\nu=0) &= \frac{4\bar{J}^2}{\beta} \sum_n \int \frac{\dd[2]{k_1}}{(2\pi)^2} \frac{\dd[2]{k_2}}{(2\pi)^2} \frac{\bigl [d_\alpha(\kk_1) d_\beta(\kk_2) + d_\beta(\kk_1) d_\alpha(\kk_2) \bigr ]\e{-i(\kk_1 - \kk_2)\cdot \Delta \rr}}{\left [(i\omega_{n} + \mu)^2 - d^2(\kk_1) \right]\left [(i\omega_{n} + \mu)^2 - d^2(\kk_2) \right]},
\end{align}
the statement $T_{\alpha z}^\text{sym}(\Delta \rr, i\omega_\nu=0)=0$ in the main text follows by replacing $\kk_1 \leftrightarrow -\kk_2$ since $d_z(\kk)$ is constant and $d_{\alpha \neq z}(-\kk) = - d_{\alpha \neq z}(\kk)$. The same momentum relabeling in one of the terms in the numerator leads to
\begin{align}
    T_{xy}^\text{sym}(\Delta \rr, i\omega_\nu=0) &=  \frac{4\bar{J}^2}{\beta} \sum_n \int \frac{\dd[2]{k_1}}{(2\pi)^2} \frac{\dd[2]{k_2}}{(2\pi)^2} \frac{2d_x(\kk_1) d_y(\kk_2) \e{-i(\kk_1 - \kk_2)\cdot \Delta \rr}}{\left [(i\omega_{n} + \mu)^2 - d^2(\kk_1) \right]\left [(i\omega_{n} + \mu)^2 - d^2(\kk_2) \right]}
\end{align}
which means that the symmetric off-diagonal spin-spin interaction is given by
\begin{align}
    T_{xy}^\text{sym}(\Delta \rr, i\omega_\nu=0) &= \frac{-8\bar{J}^2}{\beta} \sum_n V_x(\Delta \rr, i\omega_n) V_y(\Delta \rr, i\omega_n).
\end{align}
Each term in the induced exchange interaction,
\begin{align}\label{eq:J_static_starting}
    J_\alpha(\Delta \rr, i\omega_\nu=0)
&= \frac{4\bar{J}^2}{\beta} \sum_n \int \frac{\dd[2]{k_1}}{(2\pi)^2} \frac{\dd[2]{k_2}}{(2\pi)^2} \frac{(i\omega_{n} + \mu)^2  + d_\alpha(\kk_1) d_\alpha(\kk_2) - \sum_{\gamma \neq \alpha} d_\gamma(\kk_1) d_\gamma(\kk_2)}{\left [(i\omega_{n} + \mu)^2 - d^2(\kk_1) \right]\left [(i\omega_{n} + \mu)^2 - d^2(\kk_2) \right]}\e{-i(\kk_1 - \kk_2)\cdot \Delta \rr},
\end{align}
is immediately recognizable as a product of the same $U,V_\gamma$ as above. Using their $\Delta \rr$ (anti-) symmetry properties, we find
\begin{align}
    J_\alpha(\Delta \rr, i\omega_\nu=0) &= \frac{4\bar{J}^2}{\beta} \sum_n \biggl \{ \left [ (i\omega_n + \mu)^2 + s_{\alpha z}m_0^2 \right] U^2(\Delta \rr, i\omega_n) - s_{\alpha x}V_x^2(\Delta \rr, i\omega_n) - s_{\alpha y}V_y^2(\Delta \rr, i\omega_n) \biggr\},
\end{align}
where $s_{\alpha \gamma} = 2\delta_{\alpha \gamma} - 1.$

Thus, all the interactions $J_\alpha, T^\text{sym}_{\alpha \beta}, D_\gamma$ can be written in terms of some function of $\sqrt{m_0^2 - (i\omega_n + \mu)^2}$, where one must choose the branch of the square root with positive real part. As we will show shortly, this leads to a branch cut along the real axis that must be taken into consideration. In general, the Matsubara sum in the interactions,
\begin{align}
    \chi_{\alpha \beta}(\Delta \rr, i\omega_\nu=0) &= \frac{1}{\beta} \sum_n g_{\alpha \beta}\left( z_n \right) =  \frac{1}{2\pi i \beta} \oint \dd{z} f(z) g_{\alpha \beta}(z),
\end{align}
where $z_n = i\omega_n + \mu$, can be expressed in terms of a contour integral using the counting function 
\begin{align}
    f(z) = \frac{-\beta}{1+\e{\beta(z-\mu)}}.
\end{align}
The contour(s) must enclose all the poles $z_n$ of $f(z)$, and similarly to Ref.\ \cite{Johnsen2025}, for $\mu > 0$ the contours become the quarter circles in the first and fourth with radii tending to infinity. There need to be two contours because of the aforementioned branch cut near the real axis, i.e., when $z = x \pm i\epsilon$ for real $x$ and $\epsilon > 0$. For such $z$, the square root present in all $\chi_{\alpha \beta} (\Delta \rr, i\omega_\nu=0)$ takes the value
\begin{align}\label{eq:square_root_branch}
    \sqrt{m_0^2 - (x \pm i\epsilon)^2} \to \begin{cases}
        \mp i \sqrt{x^2 - m_0^2}, &x^2 > m_0^2\\
        \sqrt{m_0^2 - x^2}, &x^2 < m_0^2
    \end{cases},
\end{align}
as $\epsilon \to 0^+.$ All the arc integrals are exponentially suppressed by the counting function $f(z),$ meaning that the general expression for the static interactions is an integral over straight lines in the $z$ plane:
\begin{align}
    \chi_{\alpha \beta}(\Delta \rr, i\omega_\nu=0) &= \frac{1}{2\pi i \beta} \left (\int_{i\infty}^{i\epsilon} \dd{z} + \int_{-i\epsilon}^{-i\infty} \dd{z} + \int_{0+i\epsilon}^{\infty + i\epsilon} \dd{z} + \int_{\infty - i\epsilon}^{0-i\epsilon}\dd{z} \right) f(z)g_{\alpha \beta}(z).
\end{align}
With the guiding principle of Eq.\ \eqref{eq:square_root_branch} and a table of integrals, the three types of interactions can be calculated separately at zero temperature, where the counting function $f$ becomes a step function.

For the DM-like interaction, the imaginary-axis integrals cancel, and there is only a nonzero contribution if $\mu > m_0$. It reduces to
\begin{align}
    D_\gamma(\Delta \rr, i\omega_\nu = 0) &= \theta(\waschanged{\mu - m_0})\frac{\bar{J}^2 \abs{m_0}^3}{\abs{\Delta \rr} \pi^3 (\hbar \vF)^4} \begin{pmatrix}
        -\Delta y\\ \Delta x\\0
    \end{pmatrix}_\gamma \Im i \Re\int_0^{\mu^2/m_0^2 -1} \dd{x} \sqrt{x} \\\nonumber
    &\phantom{QQQQ} \times K_0\left(i \frac{\abs{\Delta \rr}\abs{m_0}}{\hbar \vF}\sqrt{x}\right)K_1\left(i \frac{\abs{\Delta \rr}\abs{m_0}}{\hbar \vF}\sqrt{x}\right)
\end{align}
at zero temperature, which can be computed analytically. Introducing 
\begin{align}
    \Delta \bar{\rr} \equiv \Delta \rr \frac{\sqrt{\mu^2 - m_0^2}}{\hbar \vF} =\Delta \rr \kF,
\end{align}
the result is
\begin{align}
    D_\gamma(\Delta \rr, i\omega_\nu=0) &= \theta(\waschanged{\mu - m_0}) \frac{\bar{J}^2 \abs{m_0}^3}{(\hbar \vF)^4}\begin{pmatrix}  -\sin \phi\\ \cos \phi \\0 \end{pmatrix}_\gamma \frac{\left(\frac{\mu^2}{m_0^2} -1\right)^{3/2}}{2\pi \abs{\Delta \bar{\rr}}} J_1\left(\abs{\Delta \bar{\rr}}\right) Y_1\left(\abs{\Delta \bar{\rr}} \right)
\end{align}
for the choice of SOC $\vb{d}_1$ and
\begin{align}
    D_\gamma(\Delta \rr, i\omega_\nu=0) &= \theta(\waschanged{\mu - m_0}) \frac{\bar{J}^2 \abs{m_0}^3}{(\hbar \vF)^4}\begin{pmatrix}  -\cos \phi\\ -\sin \phi \\0 \end{pmatrix}_\gamma \frac{\left(\frac{\mu^2}{m_0^2} -1\right)^{3/2}}{2\pi \abs{\Delta \bar{\rr}}} J_1\left(\abs{\Delta \bar{\rr}}\right) Y_1\left(\abs{\Delta \bar{\rr}} \right)
    \label{eq:D_gamma}
\end{align}
for the other choice. $J_n, Y_n$ are Bessel functions of the first and second kind, respectively. The large-distance behavior mentioned in the main text comes from the expansion of
\begin{align}
    \frac{J_1(x) Y_1(x)}{x} \to \frac{\cos(2x)}{\pi x^2}
\end{align}
to second order in $1/x$ for large values of $x$. Note the $p$-wave angular dependence of $D_x, D_y$.

We lack analytic expressions for the remaining interaction types, where not even the imaginary-axis integrals cancel. The result at zero temperature and $\mu>m_0$ is 
\begin{align}    
    T_{xy}^\text{sym}(\Delta \rr, i\omega_\nu=0) &=\pm \frac{\bar{J}^2 \abs{m_0}^3}{(\hbar \vF)^4} \sin(2\phi) \frac{1}{\pi^3} \left [ \Im  \int_{1}^{\mu/\abs{m_0}}\dd{u} \left( u^2-1\right) K_1^2\left (i\abs{\Delta \bar{\rr}}\frac{\sqrt{u^2-1}}{\sqrt{\frac{\mu^2}{m_0^2} - 1}}\right) \right. \\\nonumber
    &\phantom{QQ} \left.- \left(\frac{\mu^2}{m_0^2} - 1\right)^{3/2}  \int_0^{\infty} \dd{u} \left(\frac{1}{\frac{\mu^2}{m_0^2} - 1} + u^2\right) K_1^2\left(\abs{\Delta \bar{\rr}} \sqrt{\frac{1}{\frac{\mu^2}{m_0^2} -1} + u^2}\right)\right]
\end{align}
for the off-diagonal part, where the upper sign is for $\vb{d}_1$ and the lower for $\vb{d}_2$. The spin-diagonal part is given by
\begin{align}
    &J_\alpha(\Delta \rr, i\omega_\nu=0) = \frac{\bar{J}^2 \abs{m_0}^3}{\pi^3 (\hbar \vF)^4} \left \{ \int_0^{\infty} \dd{\tilde{y}} \left [ \left(-\tilde{y}^2 + s_{\alpha z} \right) K_0^2\left( \frac{\abs{\Delta \bar{\rr}}}{\sqrt{\frac{\mu^2}{m_0^2} - 1}}\sqrt{\tilde{y}^2 + 1} \right)  \right.\right. \\\nonumber
     & +  \left.\left( \tilde{y}^2 + 1 \right) K_1^2\left( \frac{\abs{\Delta \bar{\rr}}}{\sqrt{\frac{\mu^2}{m_0^2} - 1}}\sqrt{\tilde{y}^2 + 1} \right) \left(s_{\alpha x}\sin^2\phi + s_{\alpha y} \cos^2\phi\right) \right ]\\ \nonumber
     &\left.- \int_{1}^{\mu/\abs{m_0}} \dd{\tilde{x}} \left [ -\left(\tilde{x}^2 + s_{\alpha z} \right) \Im K_0^2 \left ( \frac{i\abs{\Delta \bar{\rr}}}{\sqrt{\frac{\mu^2}{m_0^2} - 1}} \sqrt{\tilde{x}^2 - 1} \right)  + \left(\tilde{x}^2 -1\right) \Im K_1^2\left ( \frac{i\abs{\Delta \bar{\rr}}}{\sqrt{\frac{\mu^2}{m_0^2} - 1}} \sqrt{\tilde{x}^2 - 1} \right)\left(s_{\alpha x}\sin^2\phi + s_{\alpha y} \cos^2\phi\right) \right] \right \},
\end{align}
where we defined
$s_{\alpha \gamma} = 2\delta_{\alpha \gamma} - 1$. The above is for $\vb{d}_1$ SOC, and the $\vb{d}_2$ choice can be obtained by interchanging $s_{\alpha x} \leftrightarrow s_{\alpha y}$. \waschanged{In the expressions and descriptions in the main text, we extend the calculations to $\mu < -m_0,$ which is done by considering the parity of each type of interactions as a function of $\mu$, using the starting expressions, Eq.~\eqref{eq:D_gamma_static_start}, \eqref{eq:T_static_starting}, and \eqref{eq:J_static_starting}.}

\section{Momentum integrals and Matsubara sums for dynamics}
\label{sec:dynamic_appendix}
This Appendix describes the general steps needed to obtain the Taylor expansion of $\chi_{\alpha \beta}(\bq, i\omega_\nu)$. Starting with Eq.\ \eqref{eq:taylor_coeff_order}, we immediately rewrite the $\kk$ integral in polar coordinates such that $k_x = k \cos\theta,$ $k_y = k \sin \theta$,
\begin{align}
    &\chi_{\alpha \beta}^{(k, i, j)} =
    \frac{1}{\beta (i!) (j!) (k!)}\int_0^\Lambda\frac{\dd{k} k}{(2\pi)^2} \int_0^{2\pi} \dd{\theta} \sum_n \lim_{\bq \to 0, i\omega_\nu \to 0} \left(\partial_{q_x}\right)^i \left(\partial_{q_y}\right)^j \left(\partial_{(i\omega_\nu)}\right)^k \chi_{\alpha \beta}(\kk+\bq, \kk, i\omega_n + i\omega_\nu, i\omega_n).
\end{align}
$\Lambda \propto 1/a$ is a momentum cutoff, appearing in our calculations explicitly for the first time here because it cannot be taken as infinity right away in this context. We omit the details of taking the above derivatives and limits of Eq.\ \eqref{eq:DTJ_general_start}, keeping the discussion general. The crucial observation needed to obtain manageable expressions is that the limit
\begin{align}
    \lim_{\bq \to 0, i\omega_\nu \to 0} \left(\partial_{q_x}\right)^i \left(\partial_{q_y}\right)^j \left(\partial_{(i\omega_\nu)}\right)^k \chi_{\alpha \beta}(\kk+\bq, \kk, i\omega_n + i\omega_\nu, i\omega_n) &= \frac{\sum_l X_{\alpha \beta, l}^{(k,i,j)}(k, i\omega_n) Y_{\alpha \beta, l}^{(k,i,j)}(\theta) }{Z_{\alpha \beta}^{(k,i,j)}(k, i\omega_n)}
\end{align}
takes the form of a finite sum of terms $X_l(k,i\omega_n) Y_l(\theta)$ divided by a completely $\theta$-independent denominator $Z(k,i\omega_n)$. It becomes $\theta$ independent because the only $\theta$ dependence in the denominator is through $\kk\cdot\bq$, but $\bq \to 0$. Thus, the $\theta$ integral can be taken inside the $n$ sum without affecting the analytic structure of the sum; the functions $Y_l(\theta)$ are prefactors. Upon integrating out $\theta$, many terms drop out due to symmetry.

The next step is to compute the $n$ sum in
\begin{align}
    &\chi_{\alpha \beta}^{(k, i, j)} =
    \frac{1}{\beta (i!) (j!) (k!)}\int_0^\Lambda\frac{\dd{k} k}{(2\pi)^2}\sum_n \frac{\sum_l X_{\alpha \beta, l}^{(k,i,j)}(k, i\omega_n) \int_0^{2\pi} \dd{\theta}  Y_{\alpha \beta, l}^{(k,i,j)}(\theta) }{Z_{\alpha \beta}^{(k,i,j)}(k, i\omega_n)},
\end{align}
which is done in the standard way, making each $n$ sum take the form
\begin{align}
    \frac{1}{\beta} \sum_n g(i\omega_n) &= \sum_{z_0\in \text{poles of } g(z)} \Res_{z_0}\Bigl [ g(z) f(z) \Bigr],
\end{align}
where $f(z)$ is the Fermi-Dirac distribution. The poles of the functions $g(z)$ in question can be of various order $N$, but they are always located at $d_\pm(\kk) = -\mu\pm \sqrt{(\hbar \vF k)^2 + m_0^2}$, so the relevant residues can be straightforwardly computed using the formula
\begin{align}
     \Res_{z_0}\Bigl [ g(z) f(z) \Bigr] &= \frac{1}{(N-1)!} \lim_{z\to z_0} \dv[N-1]{}{z} \left[(z-z_0)^N f(z) g(z)\right].
\end{align}
Here, the poles are of order $N=4$ or less, meaning our Taylor coefficient $\chi_{\alpha \beta}^{(k, i, j)}$ will contain first, second and third derivatives of the Fermi-Dirac distribution. At this stage, it is convenient to take the zero-temperature limit in order to get step functions in $k$ instead of Fermi-Dirac distributions. The presence of derivatives of these nonanalytic functions may seem problematic, but many of them turn out to vanish in the zero-temperature limit. We start with the third derivatives. The third-derivative terms for our system always take on the form
\begin{align}
    \int \dd{k} k \frac{a_0 + a_2 k^2 + a_4 k^4 + a_6 k^6}{\left[(\hbar \vF k)^2 + m_0^2 \right]^p} f'''(d_\pm(\kk))
\end{align}
for some positive (half) integer $p$, and $k$-independent coefficients $a_i$. Here, the primes mean 
\begin{align}
    f'''(d_\pm(\kk)) = \left. \left [ \dv[3]{}{z} \frac{1}{1+\e{\beta z}} \right] \right|_{z=d_\pm(\kk)},
\end{align}
but this can be related to the direct $k$ derivative of $f(d_\pm(\kk))$ through
\begin{align}
    \dv{f''(d_\pm(\kk))}{k} &= \dv{d_\pm(\kk)}{k} f'''(d_\pm(\kk)).
\end{align}
The derivative of the dispersion relation $d_\pm(\kk)$ is zero only at $k=0$, where its first-order Taylor expansion is $\pm (\hbar \vF)^2 k / m_0$. Because of the existing $k$ factor in the integrand, this zero will not be a problem, and
\begin{align}
    \int \dd{k} k \frac{a_0 + a_2 k^2 + a_4 k^4 + a_6 k^6}{\left[(\hbar \vF k)^2 + m_0^2 \right]^p} f'''(d_\pm(\kk)) &= \int \dd{k} k \frac{a_0 + a_2 k^2 + a_4 k^4 + a_6 k^6}{\left[(\hbar \vF k)^2 + m_0^2 \right]^p} \frac{\dv{f''(d_\pm(\kk))}{k}}{\dv{d_\pm(\kk)}{k}}
\end{align}
where we insert $d_\pm(\kk) = -\mu \pm \sqrt{(\hbar \vF k)^2 + m_0^2}$ to find
\begin{align}
&\int \dd{k} k \frac{a_0 + a_2 k^2 + a_4 k^4 + a_6 k^6}{\left[(\hbar \vF k)^2 + m_0^2 \right]^p} f'''(d_\pm(\kk)) = \frac{\pm 1}{(\hbar \vF)^2} \int \dd{k} \frac{a_0 + a_2 k^2 + a_4 k^4 + a_6 k^6}{\left[(\hbar \vF k)^2 + m_0^2 \right]^{p-1/2}}\dv{f''(d_\pm(\kk))}{k} = 0
\end{align}
in the $T=0$ limit since the $k$ derivative can be moved to the other factor in the integrand using integration by parts, yielding integrands proportional to the second derviative of $f(z)$, which is zero:
\begin{align}
    f''(z) &= \frac{\beta ^2 e^{\beta  z} \left(e^{\beta  z}-1\right)}{\left(e^{\beta  z}+1\right)^3} \to 0
\end{align}
for all $z\in \mathbb{R}$ in the $\beta \to \infty$ limit. We are thus rid of both second and third derivatives of $f$. We treat the first derivatives of $f(z)$ as Dirac-delta functions in the zero-temperature limit and linearize $d_\pm(\kk)$ around its zeros to obtain
\begin{align}
    f'(d_+(\kk)) &\to \begin{cases}
        -\frac{\delta \left( k - \frac{\sqrt{\mu^2 - m_0^2}}{\hbar \vF} \right)}{\hbar \vF \sqrt{1 - \frac{m_0^2}{\mu^2}}} & \text{ if } \mu > m_0\\
        0 & \text{ otherwise}
    \end{cases}\\
    f'(d_-(\kk)) &\to \begin{cases}
        -\frac{\delta \left( k - \frac{\sqrt{\mu^2 - m_0^2}}{\hbar \vF} \right)}{\hbar \vF \sqrt{1 - \frac{m_0^2}{\mu^2}}} & \text{ if } \mu < - m_0\\
        0 & \text{ otherwise}
    \end{cases}.
\end{align}
Splitting into the two cases of $\mu$ inside and outside the gap $m_0$, the integrals over $k$ can be computed for each $i,j,k, \alpha, \beta$.

Having explained the general methodology, we now print the results of these long calculations. They are
\begin{align}
    J_x(\bq, i\omega_\nu, \abs{\mu} > m_0) &= \bar{J}^2 \left[\frac{ m_0^2 }{12 \pi \abs{\mu}^3 (\hbar v_{\text{F}})^2} (i\omega _{\nu })^2 
        +\frac{\sgn\mu\left(\mu ^2 + m_0^2\right)}{4 \pi  \mu^2 (\hbar v_{\text{F}})^2}i\omega _{\nu }
        -\frac{\Lambda }{2 \pi  \hbar  v_{\text{F}}}\right.\\\nonumber
    &\phantom{QQQQ}\left.+\frac{ m_0^2\left(\mu ^2-3 m_0^2\right) }{4 \pi  \abs{\mu}^5} q_x^2
    -\frac{m_0^2 \left(\mu ^2+3 m_0^2\right)}{12 \pi  \abs{\mu}^5} q_y^2  \right]\\
    J_z(\bq, i\omega_\nu, \abs{\mu} > m_0) &= \bar{J}^2 \left [-\frac{ m_0^2 }{6 \pi \abs{\mu}^3 (\hbar v_{\text{F}})^2} (i\omega _{\nu })^2
        +\frac{\sgn\mu\left(\mu ^2-m_0^2\right) }{2 \pi  \mu ^2 (\hbar v_{\text{F}})^2}i \omega_{\nu }
        +\frac{\abs{\mu}}{\pi (\hbar \vF)^2}
        -\frac{\Lambda }{2 \pi  \hbar  v_{\text{F}}} \right. \\\nonumber
   &\phantom{QQQQ} \left.  +\frac{m_0^2(12 m_0^2-7 \mu ^2) }{12 \pi  \abs{\mu} ^5} q_x^2
    +\frac{m_0^2(12 m_0^2-7 \mu ^2) }{12 \pi  \abs{\mu} ^5} q_y^2 \right]
\end{align}
for $J_\alpha$ and $\mu$ outside the gap. $J_y$ can be obtained from $J_x$ by letting $q_x \leftrightarrow q_y$. The symmetric off-diagonal part is
\begin{align}
    T_{xy}(\bq, i\omega_\nu, \abs{\mu} > m_0) &= \bar{J}^2 \frac{m_0^2\left(3 m_0^2-2 \mu ^2 \right) }{6 \pi  \abs{\mu}^5} q_x q_y\\
    T_{xz}(\bq, i\omega_\nu, \abs{\mu} > m_0) &= \bar{J}^2\left[\frac{m_0\left(2 m_0^2-\mu ^2 \right)\sgn \mu }{4 \pi  \mu ^4 \hbar  v_{\text{F}}}i \omega _{\nu }  q_y+ \frac{m_0^3}{2 \pi  \abs{\mu}^3 \hbar  v_{\text{F}}} q_y \right] \\
    T_{yz}(\bq, i\omega_\nu, \abs{\mu} > m_0) &=\bar{J}^2 \left [-\frac{m_0\left(2 m_0^2-\mu ^2 \right)\sgn \mu }{4 \pi  \mu ^4 \hbar  v_{\text{F}}}i \omega _{\nu } q_x  - \frac{m_0^3}{2 \pi  \abs{\mu}^3 \hbar  v_{\text{F}}} q_x \right]
\end{align}
and finally 
\begin{align}
    D_x(\bq, i\omega_\nu, \abs{\mu} > m_0) &= i\bar{J}^2 \left(\frac{m_0^2 }{4 \pi  \abs{\mu}^3 \hbar  v_{\text{F}}} i\omega_\nu q_y - \frac{\sgn \mu}{2 \pi  \hbar  v_{\text{F}}} q_y\right)\\
    D_y(\bq, i\omega_\nu, \abs{\mu} > m_0) &= i\bar{J}^2\left(-\frac{m_0^2 }{4 \pi \abs{\mu}^3 \hbar v_{\text{F}}} i\omega_\nu q_x + \frac{\sgn \mu}{2 \pi  \hbar  v_{\text{F}}} q_x\right)\\
    D_z(\bq, i\omega_\nu, \abs{\mu} > m_0) &= -i \bar{J}^2\frac{ m_0 \sgn \mu}{4 \pi  \mu ^2 (\hbar v_{\text{F}})^2}(i\omega _{\nu })^2
\end{align}
for $\abs{\mu} > m_0$. Inside the gap $|\mu| < m_0$, we find that all the exchange interactions are $\mu$ independent and given by
\begin{align}
    \label{eq:T_coeffs_in_gap}
    J_x(\bq, i\omega_\nu, \abs{\mu} <m_0) &= \bar{J}^2 \left [-\frac{\Lambda }{2 \pi  \hbar \vF} -\frac{(i\omega _{\nu })^2}{6 \pi  m_0 (\hbar \vF)^2} + \frac{q_x^2}{6 \pi  m_0}\right]\\
    J_y(\bq, i\omega_\nu, \abs{\mu} <m_0) &= \bar{J}^2 \left [-\frac{\Lambda }{2 \pi  \hbar \vF} - \frac{(i\omega _{\nu })^2}{6 \pi  m_0 (\hbar \vF)^2}+ \frac{q_y^2}{6 \pi  m_0}\right]\\
    J_z (\bq, i\omega_\nu,\abs{\mu} <m_0) &= \bar{J}^2 \left [-\frac{\Lambda }{2 \pi  \hbar \vF} -\frac{(i\omega _{\nu })^2}{6 \pi  m_0 (\hbar  v_{\text{F}})^2} + \frac{2m_0}{ \pi  (\hbar \vF)^2} + \frac{q_x^2}{6 \pi  m_0} + \frac{q_y^2}{6 \pi  m_0} \right ]
\end{align}
and similarly for the DM interactions
\begin{align}
    D_{\gamma \neq z}(\bq, i\omega_\nu, \abs{\mu} < m_0) &= 0\\
    D_z(\bq, i\omega_\nu, \abs{\mu} < m_0) &=\frac{i \bar{J}^2}{2 \pi (\hbar v_{\text{F}})^2 } i\omega_{\nu}
\end{align}
and for the off-diagonal exchange interactions
\begin{align}
    T_{xy}(\bq, i\omega_\nu, \abs{\mu} < m_0) &= \bar{J}^2\frac{ q_x q_y}{6 \pi  m_0}\\
    T_{xz}(\bq, i\omega_\nu, \abs{\mu} < m_0) &=0\\
    T_{yz}(\bq, i\omega_\nu, \abs{\mu} < m_0) &= 0.
\end{align}
The cutoff-dependent terms in $J_\alpha$ are $\alpha$ independent, so in the action they will, according to Eq.\ \eqref{eq:lagrangian_for_eom}, appear as a constant term $\propto \Lambda \nn(\rr, t) \cdot \nn(\rr, t)$ since $\nn \cdot \nn$ is constant \cite{Nogueira2018}. We note that they have the same value as the equivalent term in Ref.\ \cite{Nogueira2018}, and that they do not affect the magnetization dynamics, where they give rise to a term $\propto \Lambda \nn \cross\nn = 0$ in the LLG equation, Eq.\ \eqref{eq:H_eff_prime_form}.

\section{Equation of motion}
\label{sec:llg_appendix}
This Appendix shows how to arrive at an equation of motion for the magnetization based on the real-time action in Eq.\ \eqref{eq:action_realtime_rt} and the Taylor expansion in Eq.\ \eqref{eq:taylor_imaginary}, where we keep terms up to second order. Explicitly,
\begin{align}
     \Delta S_\text{eff}^{(\text{2, RT})} &= \int\dd[2]{\rr_1} \dd[2]{\rr_2} \int\frac{\dd t_1 \dd t_2}{2\pi} \int\frac{\dd \omega}{2\pi} \int \frac{\dd[2]{q}}{(2\pi)^2} n_\alpha(\rr_1, t_1) \e{-i(\bq\cdot \rr_1 - \omega t_1)} \biggl [\chi_{\alpha \beta}^{(0,0)} + \chi_{\alpha \beta \gamma}^{(0,1)} q_\gamma \\\nonumber
     &\phantom{QQQQQQ}+ \chi_{\alpha \beta}^{(1,0)} \omega + \chi_{\alpha \beta \gamma}^{(1,1)}\omega q_\gamma + \chi_{\alpha \beta \gamma \delta}^{(0,2)} q_\gamma q_\delta + \chi_{\alpha \beta}^{(2,0)} \omega^2 \biggr ] \e{i(\bq \cdot \rr_2 - \omega t_2)} n_\beta(\rr_2, t_2)
\end{align}
where the replacements $\omega \to \partial_{t_2}/(-i)$, $q_\gamma \to \partial_{r_{2,\gamma}}/i$ can be made, such that the differential operators act on the exponential function on the right. Next we integrate by parts as many times as needed to move the derivatives to $n_\beta$,
\begin{align}
    \Delta S_\text{eff}^{(\text{2, RT})} &= \int\dd[2]{\rr_1} \dd[2]{\rr_2} \int\frac{\dd t_1 \dd t_2}{2\pi} \int\frac{\dd \omega}{2\pi} \int \frac{\dd[2]{q}}{(2\pi)^2} n_\alpha(\rr_1, t_1) \e{i \bigl [\bq \cdot(\rr_2 - \rr_1) - \omega(t_2 - t_1)\bigr]} \biggl [ \chi_{\alpha \beta}^{(0,0)} \\\nonumber
    &\phantom{QQQ} - \chi_{\alpha \beta \gamma}^{(0,1)} \frac{1}{i} \partial_{r_{2,\gamma}} - \chi_{\alpha \beta }^{(1,0)} \frac{1}{-i} \partial_{t_2} + \chi_{\alpha \beta \gamma}^{(1,1)}\frac{1}{-i}\partial_{t_2} \frac{1}{i}\partial_{r_{2,\gamma}} + \chi_{\alpha \beta \gamma \delta}^{(0,2)} \frac{1}{(i)(i)}\partial_{r_{2,\gamma}} \partial_{r_{2,\delta}}  + \chi_{\alpha \beta}^{(2,0)} \frac{1}{i^2}\partial_{t_2}^2 \biggr ] n_\beta(\rr_2, t_2),
\end{align}
where the $\omega$ and $\bq$ integrals yield Dirac-delta functions in space and time, immediately yielding Eq.\ \eqref{eq:lagrangian_for_eom} in the main text. We omit the details here, but to derive the proper equation of motion, one needs to take into account the Berry phase term $\vb{b}\cdot \dot{\nn}$ in $\mathcal{L}_\text{FMI}$, where the Berry connection $\vb{b}$ satisfies $\partial_\nn \cross \vb{b} = -\nn/\nn^2$. Then, the equations of motion stemming from the Euler-Lagrange equations take the form of an LLG equation,
\begin{align}
    \dot{\nn} &= \nn \cross \left ( \tilde{\vb{L}} + \vb{L}^{(2)} \right),
\end{align}
where $\tilde{\vb{L}}$ is the contribution from all other terms than $\Delta S_\text{eff}^{(2)}$ and the Berry-phase term in Eq.\ \eqref{eq:action_start}. In the same way the induced interactions were treated, it can be shown that the FM's intrinsic interactions, shown in Eq.\ \eqref{eq:FM_hamiltonian}, lead to 
\begin{align}
    \tilde{\vb{L}} = -\frac{K}{\pi} \left( \tilde{m}_0 + n_z \right)\hat{\vb{z}} -\frac{J_\text{ex}}{\pi} (\partial_x^2 + \partial_y^2) \nn.
\end{align}
The term we are interested in, $\vb{L}^{(2)}$, comes from applying the Euler-Lagrange equations to $\Delta S_\text{eff}^{(2)}$, i.e.,
\begin{align}
    L_i^{(2)} &=
    \pdv{\mathcal{L}_\text{eff}^{(2)}}{n_i} 
    - \partial_t \pdv{\mathcal{L}_\text{eff}^{(2)}}{\dot{n}_i} 
    - \partial_\mu \pdv{\mathcal{L}_\text{eff}^{(2)}}{(\partial_\mu n_i)} 
    + \partial_t^2 \pdv{\mathcal{L}_\text{eff}^{(2)}}{\ddot{n}_i} 
    + \partial_\mu^2 \pdv{\mathcal{L}_\text{eff}^{(2)}}{(\partial_\mu^2 n_i)}  + \partial_t \partial_\mu \pdv{\mathcal{L}_\text{eff}^{(2)}}{(\partial_\mu \dot{n}_i)} 
    + \partial_x\partial_y \pdv{\mathcal{L}_\text{eff}^{(2)}}{(\partial_x \partial_y n_i)},
\end{align}
where summation over repeated indices $\mu=x,y$ is implied. The first Euler-Lagrange term gets a contribution from all terms in the Lagrangian, 
\begin{align}
    2\pi \pdv{\mathcal{L}_\text{eff}^{(2)}}{n_i}  &= \biggl ( \chi_{i \beta}^{(0,0)} + \chi_{\beta i }^{(0,0)} + i \chi_{i \beta \gamma}^{(0,1)} \partial_{\gamma} - i \chi_{i \beta }^{(1,0)}  \partial_{t} + \chi_{i \beta \gamma}^{(1,1)}\partial_{t} \partial_{\gamma} - \chi_{i \beta \gamma \delta}^{(0,2)} \partial_{\gamma} \partial_{\delta} - \chi_{i \beta}^{(2,0)} \partial_{t}^2 \biggr ) n_\beta(\rr, t),
\end{align}
and, for example, the penultimate Euler-Lagrange term gets a contribution
\begin{align}
    2\pi \partial_t \partial_\mu \pdv{\mathcal{L}_\text{eff}^{(2)}}{(\partial_\mu \dot{n}_i)} &= \partial_t \partial_\mu \chi_{\alpha i \mu}^{(1,1)} n_\alpha = \chi_{\beta i \gamma}^{(1,1)} \partial_t \partial_\gamma n_\beta,
\end{align}
which has interchanged $i \leftrightarrow \beta$, adding to the $\partial_t\partial_y$ term in $\partial_{n_i}\mathcal{L}^{(2)}_\text{eff}$. The result of these straightforward calculations is that the $\vb{L}^{(2)}$ vector in the LLG equation has the general structure
\begin{align}
     L^{(2)}_\alpha &= \frac{1}{2\pi} \biggl [ \left( \chi_{\alpha \beta}^{(0,0)} + \chi_{\beta \alpha }^{(0,0)} \right)
     + i \left(\chi_{\alpha \beta \gamma}^{(0,1)} - \chi_{\beta \alpha \gamma}^{(0,1)} \right)\partial_{\gamma}
     - i \left( \chi_{\alpha \beta }^{(1,0)} - \chi_{\beta \alpha }^{(1,0)} \right) \partial_{t} \\\nonumber
     &+ \left(\chi_{\alpha \beta \gamma}^{(1,1)} + \chi_{\beta \alpha \gamma}^{(1,1)}\right)\partial_{t} \partial_{\gamma}
     - \left ( \chi_{\alpha \beta \gamma \delta}^{(0,2)} + \chi_{\beta \alpha \gamma \delta}^{(0,2)} \right) \partial_{\gamma} \partial_{\delta}
     - \left( \chi_{\alpha \beta}^{(2,0)} + \chi_{\beta \alpha}^{(2,0)} \right) \partial_{t}^2 \biggr ] n_\beta(\rr, t),
\end{align}
showing that certain terms present at the action level are actually irrelevant for the magnetization dynamics as derived using the Euler-Lagrange equations. These are $J_{\alpha}^{(1,0)}$, which would have been dissipation terms similar to Gilbert damping; the spin-symmetric gradient terms $T_{\alpha z, \gamma}^{(0,1)}$; the spin-antisymmetric nutation term $D_z^{(2,0)}$; and the spin-antisymmetric mixed-derivative term $D_{\gamma \neq z}^{(1,1)}$. Inserting only the nonzero coefficients, read off of Eq.s \eqref{eq:J_alpha_expansion} to \eqref{eq:D_gamma_expansion}, into the above and summing over $\beta,\gamma,\delta$ is somewhat tedious but straightforward and yields
\begin{align}
    \vb{L}^{(2)}&= \frac{1}{\pi} \left[ 
    \begin{pmatrix}
        J_x^{(0,0)} n_x\\
        J_y^{(0,0)} n_y \\
        J_z^{(0,0)} n_z
    \end{pmatrix} - T_{xy,xy}^{(0,2)}\begin{pmatrix}
        \partial_{xy} n_y\\
        \partial_{xy} n_x\\
        0
    \end{pmatrix}
     - \begin{pmatrix}
        J_x^{(2,0)} \ddot{n}_x \\
        J_y^{(2,0)} \ddot{n}_y\\
        J_z^{(2,0)} \ddot{n}_z
    \end{pmatrix} - \begin{pmatrix}
        \left ( J_{x, xx}^{(0,2)}\partial_x^2 +J_{x, yy}^{(0,2)}\partial_y^2 \right) n_x\\
        \left ( J_{y, xx}^{(0,2)}\partial_x^2 +J_{y, yy}^{(0,2)}\partial_y^2 \right) n_y\\
        \left ( J_{z, xx}^{(0,2)}\partial_x^2 +J_{z, yy}^{(0,2)}\partial_y^2 \right) n_z
    \end{pmatrix} \right. \\ \nonumber
    &\phantom{QQQQQQ} \left.+ \begin{pmatrix}
        T^{(1,1)}_{xz,y}  \partial_y \dot{n}_{z}\\
        T^{(1,1)}_{yz,x}  \partial_x \dot{n}_{z}\\
        T^{(1,1)}_{zx,y}  \partial_y \dot{n}_{x} + T^{(1,1)}_{zy,x}  \partial_x \dot{n}_{y}
    \end{pmatrix}
    + i \begin{pmatrix}
        -D_{y,x}^{(0,1)} \partial_x n_z\\
        D_{x,y}^{(0,1)} \partial_y n_z\\
        -D_{x,y}^{(0,1)} \partial_y n_y + D_{y,x}^{(0,1)}\partial_x n_x
    \end{pmatrix}  +  i D^{(1,0)} \hat{\vb{z}} \cross \dot{\nn}\right].
\end{align} 
where the coefficients $D_{\gamma, \delta}^{(0,1)},D_{\gamma}^{(1,0)}$ are purely imaginary and the rest are real, making $\vb{L}^{(2)}$ real. Several of the interaction coefficients above are equal (sometimes up to a sign) as can be seen in Appendix \ref{sec:dynamic_appendix}, which is taken into account in Eq.\ \eqref{eq:H_eff_prime_form} from the main text along with an absorption of the factor $\pi$ to form the primed quantities.

Although a general solution of the equation of motion is not available, some analytical progress can be made to understand the effect of the new terms present by considering linear spin waves with small amplitudes. This involves inserting the Ansatz
\begin{align}
     \nn(\rr, t) = \tilde{m}_0 \hat{\vb{z}} + \epsilon \vb{s}\e{i(\kk \cdot \rr - \omega t)}
\end{align}
into the LLG equation and linearizing it, i.e., keeping terms only up to first order in the smallness parameter $\epsilon.$ Here, $\epsilon \vb{s}= \epsilon(s_x,s_y,s_z)$ is a vector with the amplitudes of the fluctuations in the full magnetization, and $\tilde{m}_0$ is the same quantity as in Eq.\ \eqref{eq:H_int}. The constraint $\nn(\rr,t)^2=\tilde{m}_0^2$ to first order in $\epsilon$ demands that $s_z=0$. To first order, then, the LLG equation reads
    $\mathbb{B} \vb{s} = 0,$
after canceling out $\epsilon$ and the phase factor. The matrix $\mathbb{B}$ is given by
\begin{align}
    \mathbb{B} &= \left(
\begin{array}{ccc}
      i\omega (1 + i D^{\prime(1,0)} \tilde{m}_0) - k_x k_y \tilde{m}_0 T^{\prime (0,2)}_{xy,xy}  & B_{12} & \tilde{m}_0(k_x \omega T^{\prime (1,1)} - D^{\prime(0,1)} k_y) \\
     B_{21} & i\omega (iD^{\prime(1,0)} \tilde{m}_0+1) + k_x k_y \tilde{m}_0 T^{\prime (0,2)}_{xy,xy} & \tilde{m}_0 (k_y \omega T^{\prime (1,1)}+D^{\prime(0,1)} k_x ) \\
     0 & 0 & i \omega  \\
\end{array}
\right),
\end{align}
where
\begin{subequations}
\begin{align}
    B_{12} &= -\tilde{m}_0 \left(\frac{K}{\pi} + J^{\prime (0,0)}_{xy}-J^{\prime (0,0)}_{z}+J^{\prime (2,0)}_{x} \omega ^2+J^{\prime(0,2)}_{x,xx} k_y^2+J^{\prime(0,2)}_{x,yy} k_x^2 + J_\text{ex}k^2\right)\\
    B_{21} &= \tilde{m}_0 \left(\frac{K}{\pi} + J^{\prime (0,0)}_{xy}-J^{\prime (0,0)}_{z}+J^{\prime (2,0)}_{x} \omega ^2+J^{\prime(0,2)}_{x,xx} k_x^2+J^{\prime(0,2)}_{x,yy} k_y^2+ J_\text{ex}k^2\right).
\end{align}
\end{subequations}
In light of the constraint $s_z=0$ and the zeros in the bottom row, only the top left $2\times2$ matrix needs to be considered. Then, $\mathbb{B}\vb{s}=0$  constitutes two equations for $s_x,s_y$ that do not contain the DMI term $D^{(0,1)}$ nor the mixed-derivative term $T^{(1,1)}$. The effect of these, then, is not present to linear order and with this spin-wave Ansatz. We will look for a solution of the linearized LLG equation of the form $s_x = s_y \e{i\phi} > 0$, where the phase difference $\phi=\pi/2$ or $\phi=-\pi/2$. Linearly combining the two amplitude equations and eliminating $s_x$, we obtain
\begin{align}
      2\omega (1 + i D^{\prime(1,0)} \tilde{m}_0) +  \sgn\phi (B_{21} - B_{12}) &= 0,
\end{align}
where the $T_{xy,xy}^{\prime(0,2)}$ terms cancelled out, leaving $\omega(\kk)$ isotropic in $\kk$. This equation is inherently limited to quadratic order in $k,\omega$, as our original spin theory. Proceeding now, the solution for $\omega(\kk)$ is
\begin{align}
    \omega_\pm(\kk) &= \frac{1}{2J_{xy}^{\prime(2,0)}} \Biggl[ -\left(1 + iD^{\prime(1,0)}\right) \sgn \phi \\
    &\pm \sqrt{\left(1 + iD^{\prime(1,0)}\right)^2 + 4 J_{xy}^{\prime(2,0)} \left( J_z^{\prime(0,0)} - J_{xy}^{\prime(0,0)} - \frac{K}{\pi} \right) - 2 J_x^{\prime (2,0)} \left(J^{\prime(0,2)}_{x,xx} + J^{\prime(0,2)}_{x,yy} + 2\frac{J_\text{ex}}{\pi} \right)k^2} \Biggr],
\end{align}
which needs to be expanded around $k=0$ only up to second order to maintain consistency with the spin theory:
\begin{align}
    \omega_\pm(\kk) &= \frac{1}{2J_{xy}^{\prime(2,0)}} \left [-\left(1+iD^{\prime(1,0)}\right) \sgn\phi \pm \left(\sqrt{\left(1+iD^{\prime(1,0)}\right)^2 + 4J_{xy}^{\prime (2,0)} \left( J_z^{\prime(0,0)} - J_{xy}^{\prime(0,0)} -\frac{K}{\pi}\right)} \vphantom{\frac{2J_{xy}^{\prime (2,0)} \left(J^{\prime(0,2)}_{x,xx} + J^{\prime(0,2)}_{x,yy} + 2 \right)}{2\sqrt{\left(1+iD^{\prime(1,0)}\right)^2 + 4J_{xy}^{\prime (2,0)} \left( J_z^{\prime(0,0)} - J_{xy}^{\prime(0,0)} \right)}}}\right. \right.\\\nonumber
    &\phantom{Q} \left.\left. - \frac{J_{xy}^{\prime (2,0)} \left(J^{\prime(0,2)}_{x,xx} + J^{\prime(0,2)}_{x,yy} + 2 \frac{J_\text{ex}}{\pi}\right)}{\sqrt{\left(1+iD^{\prime(1,0)}\right)^2 + 4J_{xy}^{\prime (2,0)} \left( J_z^{\prime(0,0)} - J_{xy}^{\prime(0,0)} -\frac{K}{\pi}\right)}}k^2\right)\right].
\end{align}
In the main-text Figure, we use $\phi = -\pi/2$, leading to Eq.\ \eqref{eq:disp_rel}.

\end{widetext}

\end{document}